%% file: main.tex
\documentclass[journal ]{new-aiaa}
\usepackage[utf8]{inputenc}
\usepackage{graphicx}
\usepackage[version=4]{mhchem}
\usepackage{longtable,tabularx}
\usepackage{textcomp}
\usepackage{xcolor}
\usepackage{siunitx}
\usepackage{array}
\usepackage{nomencl}
\usepackage{mathrsfs} 
\usepackage{optidef}
\usepackage{multicol}
\usepackage{svg}
\usepackage{lipsum}
\usepackage{calrsfs}
\usepackage{scalerel}
\usepackage{stackengine}
\usepackage{booktabs}
\usepackage{caption}
\usepackage{subcaption}
\usepackage{arydshln}
\usepackage{mathtools}
\usepackage[makeroom]{cancel}
\usepackage{stackengine}
\usepackage{textcase}
\usepackage[english]{babel}
\usepackage{algorithm}
\usepackage{algpseudocode}
\usepackage{amsmath}
\usepackage{newtxmath}
\usepackage{cases}        

\usepackage{etoolbox}

\makeatletter
\newcommand{\unmarkedtitlefootnote}[1]{%
  \begingroup
  \renewcommand{\thefootnote}{}
  \renewcommand{\@makefnmark}{}
  \footnotetext{#1}%
  \endgroup
}

\pretocmd{\maketitle}{%
  \unmarkedtitlefootnote{Presented as Paper 2025-2265 at the AIAA SciTech 2025 Forum, Orlando, FL, January 6--10, 2025.}%
}{}{\errmessage{Patching \string\maketitle\space failed}}
\makeatother

\usepackage{amsmath,amssymb,amsfonts}
\usepackage{bm}           

\DeclareMathOperator*{\argmin}{arg\,min}

\usepackage{comment}

\title{Control Allocation Algorithm for Hypersonic Glide Vehicles with Input Limitations}

\author{Johannes Autenrieb\footnote{Research Scientist, Department of Flight Dynamics and Simulation; johannes.autenrieb@dlr.de (Corresponding Author).}}
\affil{German Aerospace Center (DLR), Institute of Flight Systems, 38108, Braunschweig, Germany}

\author{Patrick Gruhn\footnote{Head of Missile Technologies Group, Department of Supersonic and Hypersonic Technologies; patrick.gruhn@dlr.de.}}
\affil{German Aerospace Center (DLR), Institute of Aerodynamics and Flow Technology, 51147, Cologne, Germany}

\begin{document}

\maketitle

\begin{abstract}
\input{Sections/Abstract}
\end{abstract}

\section*{Nomenclature}
\input{Sections/Nomenclature}

\section{Introduction}
\label{Introduction}
\input{Sections/Introduction}
\section{The General Control Allocation Problem}
\label{The General Control Allocation Problem}
\input{Sections/Basics_Control_Allocation}
\section{Problem Formulation}
\label{Problem Formulation}
\input{Sections/Problem_Formulation}
\section{The Integrated Overall Control Allocation Concept}
\input{Sections/Control_Allocation_Concept}

\label{The Overall Control Allocation Concept}
\section{The Proposed Online Control Allocation Algorithm}
\label{ProposedOnlineControlAllocationAlgorithm}
\input{Sections/ProposedOnlineControlAllocationAlgorithm}
\section{Simulation Results}
\label{Simulation_Results}
\input{Sections/Simulation_Results}
\section{Conclusion}
\input{Sections/Conclusions}

\bibliography{./Bibliography/references} 

\end{document}

%% file: Sections/Abstract.tex
Hypersonic glide vehicles (HGVs) operate in challenging flight regimes characterized by strong nonlinearities in actuation and stringent physical constraints. These include state-dependent actuator limitations, asymmetric control bounds, and thermal loads that vary with maneuvering conditions. This paper introduces an iterative control allocation method to address these challenges in real time. The proposed algorithm searches for control inputs that achieve the desired moment commands while respecting constraints on input magnitude and rate. For slender HGV configurations, thermal loads and drag generation are strongly correlated-lower drag typically results in reduced surface heating. By embedding drag-sensitive soft constraints, the method improves energy efficiency and implicitly reduces surface temperatures, lowering the vehicle's infrared signature. These features are particularly advantageous for long-range military operations that require low observability. The approach is demonstrated using the DLR's Generic Hypersonic Glide Vehicle 2 (GHGV-2) simulation model. The results confirm the method's effectiveness in maintaining control authority under realistic, constrained flight conditions.

%% file: Sections/Nomenclature.tex
\noindent\textbf{Symbols}
{\renewcommand\arraystretch{1.0}
\noindent\begin{longtable*}{@{}l @{\quad=\quad} l@{}}

$\mathbf{B}$                 & Control effectiveness matrix, $\mathbb{R}^{3 \times 4}$ \\
$\mathbf{B}_{\nu_x}, \mathbf{B}_{\nu_y}, \mathbf{B}_{\nu_z}$ & Conditionalized rows of $\mathbf{B}$ associated with roll, pitch, and yaw \\
$C_D$               & Aerodynamic drag coefficient \\
$\mathbf{D}$        & Attainable moment set, $\mathbb{R}^3$ \\
$J$                 & Allocation cost, $J = \|\mathbf{u}\|_2^2$ \\
$\mathbf{L} = [L,\,M,\,N]^\mathrm{T}$ & Body-axis moment vector (roll, pitch, yaw) in Nm \\
$N_{\text{iter}}$   & Number of iterations in IDCA loop \\
$\mathbf{P}$        & Cost matrix in quadratic program \\
$q$                 & Dynamic pressure in N/m$^2$ \\
$r$                 & Relative degree of the output \\
$T$                 & Time step in s \\
$\mathbf{u}$        & Control input vector, $\mathbb{R}^{4}$ \\
$u_i$               & Control input for effector $i$ in deg \\
$\dot{u}_i$         & Rate of control input $i$ in deg/s \\
$u_{\text{max},i}$  & Upper magnitude bound for $u_i$ in deg \\
$u_{\text{min},i}$  & Lower magnitude bound for $u_i$ in deg \\
$\dot{u}_{\text{max},i}$ & Upper rate bound for $u_i$ in deg/s \\
$\dot{u}_{\text{min},i}$ & Lower rate bound for $u_i$ in deg/s \\
$\mathbf{u}_s$      & Desired steady-state control input vector \\
$\mathbf{u}_r$      & Baseline control input from trajectory planning \\
$\Delta \mathbf{u}$ & Control input correction vector \\
$\boldsymbol{\nu}$  & Virtual control input vector, $\mathbb{R}^3$ \\
$\nu_x, \nu_y, \nu_z$ & Components of $\boldsymbol{\nu}$, corresponding to roll ($L$), pitch ($M$), and yaw ($N$) \\
$\Delta \boldsymbol{\nu}$ & Corrective virtual control command \\
$\mathbf{x}$     & State vector, $\mathbb{R}^n$ \\
$\mathbf{y}$     & Output vector, $\mathbb{R}^p$ \\
$\mathbf{h}(\mathbf{x})$ & Output mapping, $\mathbb{R}^n \rightarrow \mathbb{R}^p$ \\
$\text{L}_{\mathbf{i}} \mathbf{h}$ & Lie derivative of function $\mathbf{h}$ along vector field $\mathbf{i}$ (e.g, $\mathbf{f},\mathbf{g}$) \\
$\alpha$            & Angle of attack in rad \\
$\beta$             & Sideslip angle in rad \\
$\gamma$            & Flight path angle in rad \\
$\rho$              & Atmospheric air density in kg/m$^3$ \\
$\Lambda(t)$        & Time-dependent modulation of input constraints \\
$\boldsymbol{\sigma}$ & Operating point vector (e.g.\ altitude, Mach number) \\
\end{longtable*}}

\noindent\textbf{Subscripts}
{\renewcommand\arraystretch{1.0}
\noindent\begin{longtable*}{@{}l @{\quad=\quad} l@{}}
$c$     & Command or control \\
$r$     & Reference or baseline input \\
$s$     & Steady-state solution \\
$i$     & Index for control surface or effector \\
$\max$  & Maximum bound \\
$\min$  & Minimum bound \\
$\text{iter}$ & Iteration index \\
\end{longtable*}}

%% file: Sections/Introduction.tex
In recent years, hypersonic glide vehicles (HGVs) have received increasing attention in academic and industrial research and development \cite{Slayer_2022}. These vehicles, designed to navigate through lower atmospheric layers at hypersonic speeds, present unique challenges that require precise control for optimal performance and guaranteed stability. Some HGV designs incorporate redundant control effectors, necessitating a control allocation system capable of distributing desired forces and moments among a redundant set of control effectors to achieve the desired vehicle response \cite{Autenrieb_2023}.

Traditional control allocation methods for overactuated aerial vehicles have evolved significantly \cite{Oppenheimer_2006}. For most aerospace applications, the pseudoinverse-based control allocation (PICA) method \cite{Bordignon_2002}, utilizing the Moore-Penrose inverse, is commonly used due to its simplicity. However, PICA does not adequately account for control input limitations. An alternative and more precise approach is to formulate the constrained control allocation problem as a quadratic programming (QPCA) problem and solve it using a QP solver \cite{Johansen_2004,Ola2004}. This method can accurately handle asymmetric input constraints and coupling effects between control effectors. However, the computational intensity of QP solvers poses significant challenges for real-time implementation on systems with limited computational resources, such as embedded systems in hypersonic reentry vehicles \cite{Burken_2001}. The high computational demand makes it difficult to guarantee real-time performance, which is essential for the fast dynamics of HGVs. To address the need for a computationally less complex solution, Bordignon introduced the redistributed pseudoinverse control allocation (RPICA) method \cite{Bordignon_1996}. RPICA iteratively approximates feasible solutions by redistributing residuals over the attainable moment set (AMS), offering a balance between computational efficiency and practicality. A scaled variant, redistributed scaled pseudoinverse control allocation (RSPICA), was later proposed to maintain the directionality of the input vector in multiple-input, multiple-output (MIMO) systems under saturation conditions \cite{Hafner_2025}. Despite advancements in computational simplicity, both RPICA and RSPICA face challenges when dealing with systems characterized by asymmetric control input limits and strong inter-directional coupling between control effectors. This issue is particularly evident in certain classes of HGVs where only positive control deflections are permissible. The inability of these methods to effectively handle asymmetric constraints and coupling effects limits their applicability in such scenarios.

\begin{figure}
\centering
\subcaptionbox{External view of the GHGV-2 \label{fig:GHGV-External}}{\includegraphics[width=0.48\textwidth]{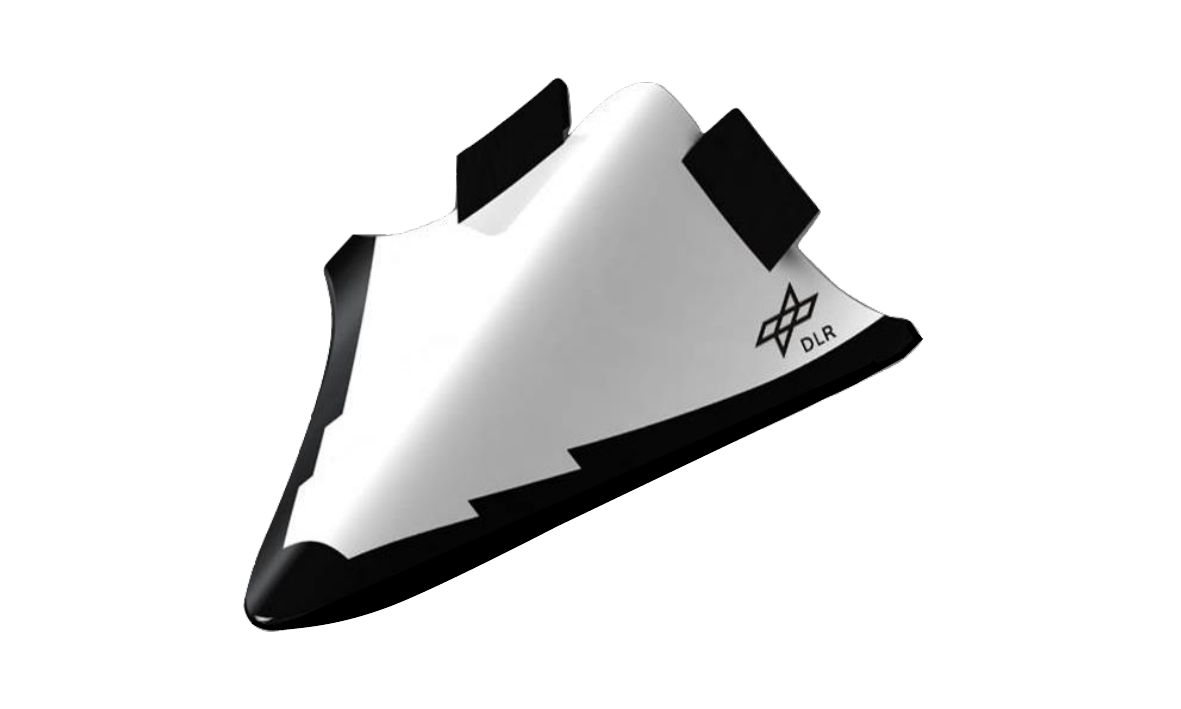}}
\hfill
\subcaptionbox{Sectional view of the GHGV-2 \label{fig:GHGV-Sectional}}{\includegraphics[width=0.48\textwidth]{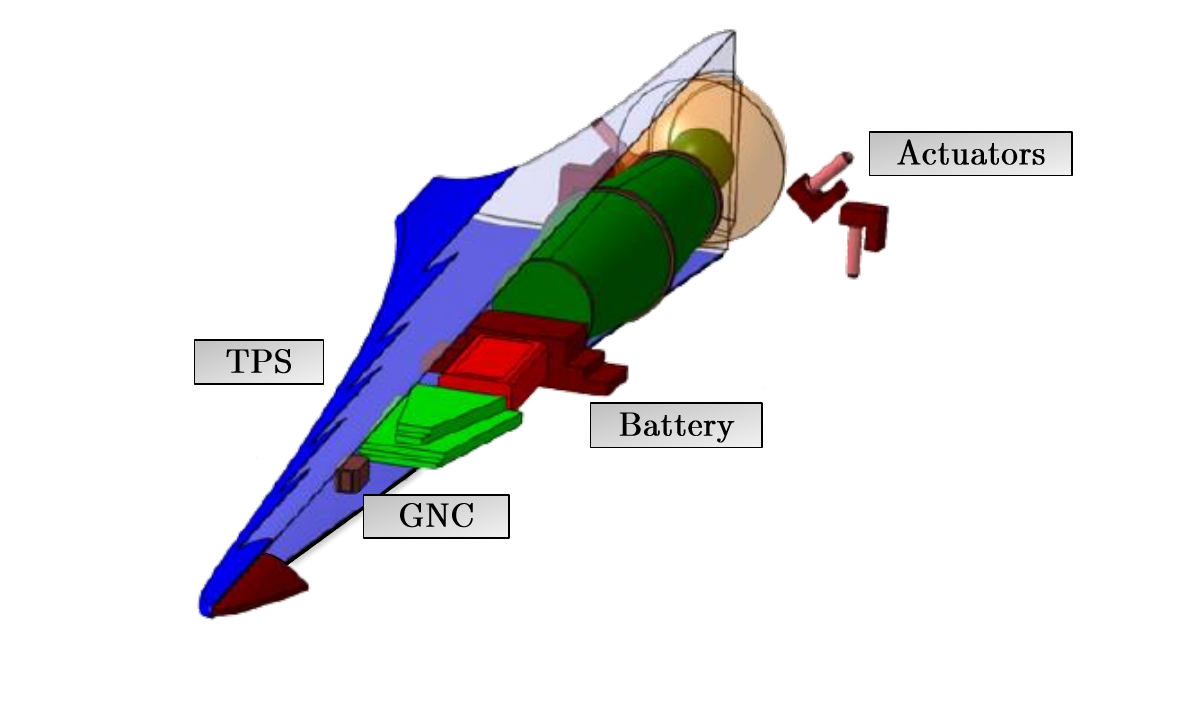}}
\caption{The DLR Generic Hypersonic Glide Vehicle 2 concept.}
\label{fig:GHGV-Concept}
\end{figure}

Over the past few years, the German Aerospace Center (DLR) has undertaken a substantial research initiative focused on the development of the Generic Hypersonic Glide Vehicle 2 (GHGV-2). This research aims to comprehensively explore the capabilities and performance characteristics of strategic hypersonic glide vehicles, shedding light on their implications for understanding hypersonic threats \cite{Autenrieb2021, Autenrieb2022}. The GHGV-2 adopts a waverider design to optimize lift-to-drag ratios at high Mach numbers \cite{Gruhn2020}. The GHGV-2 is depicted in Figure~\ref{fig:GHGV-Concept}, with an external view shown in Figure~\ref{fig:GHGV-External} and a sectional view highlighting relevant subsystems in Figure~\ref{fig:GHGV-Sectional}. Key subsystems include the thermal protection system (TPS), guidance, navigation, and control (GNC) system, battery, and actuators. During endoatmospheric flight, the GHGV-2 utilizes four integrated flaps as control effectors to regulate its attitude. Small propulsors are integrated in exoatmospheric conditions to generate the necessary moments for attitude adjustments.

The GHGV-2 is launched via a multi-stage rocket booster as part of the investigated mission profile. After ignition and acceleration, the vehicle separates from the launch system at an altitude of approximately 100\,km, initiating a parabolic re-entry trajectory. Once sufficient dynamic pressure is achieved during descent, the vehicle transitions to aerodynamic control using four surfaces,two located on the upper and two on the lower side. Following re-entry, the GHGV-2 tracks a predefined flight path angle to sustain its maximum lift-to-drag ratio en route to the target location. Due to sustained hypersonic speeds, the vehicle is exposed to significant thermal loads during this atmospheric glide phase. Consequently, any viable control allocation algorithm must generate the necessary control moments for maneuvering and account for thermal effects, ideally minimizing heat loads to reduce the burden on the TPS and lower the infrared signature, a key consideration for military applications. Another critical challenge arises from state-dependent control input limitations stemming from the wide range of operating conditions. Increased dynamic pressure at lower altitudes and higher speeds restricts actuator limits compared to flight conditions involving lower speeds or air densities. This variability demands a control allocation approach explicitly incorporating state-dependent constraints to ensure robust actuator performance throughout the mission.

Given these complexities, developing a control allocation algorithm that is both computationally efficient for real-time implementation on embedded systems and capable of handling asymmetric control input limitations, actuator coupling, thermal load considerations, and state-dependent constraints is a significant challenge. This work presents an iterative control allocation approach explicitly tailored for hypersonic glide vehicles like the GHGV-2. We begin with introducing the general control allocation problem of overactuated systems and then discuss the specific control allocation problem for HGVs, highlighting the unique challenges posed by asymmetric control input limitations, significant coupling between control effectors, thermal load considerations, and state-dependent actuator constraints due to varying flight conditions. Based on the described problem formulation, we present our proposed method that addresses these challenges by directly incorporating state-dependent control input limits and thermal load considerations into the control allocation algorithm. It iteratively adjusts the control inputs to ensure the required control moments are achieved while respecting all constraints and minimizing thermal stress on the control surfaces. The algorithm is designed to be computationally efficient, making it suitable for real-time implementation on the vehicle's embedded systems. We evaluate the effectiveness of the proposed approach through numerical simulations, demonstrating its advantages in handling the complex requirements of hypersonic glide vehicle control allocation. The results show that our method provides a reliable solution that balances computational efficiency with the ability to manage asymmetric constraints, actuator coupling, and thermal considerations.

%% file: Sections/Basics_Control_Allocation.tex
We define the following nonlinear control-affine system
\begin{equation}
\label{NonlinearSystem_1}
\dot{\mathbf x}(t) = \mathbf f(\mathbf x(t)) + \mathbf g(\mathbf x(t)) \mathbf u(t),
\end{equation}
\begin{equation}
\label{NonlinearSystem_2}
\mathbf y(t)= \mathbf h(\mathbf x(t)),
\end{equation}
where $\mathbf x(t) \in \mathbb{R}^n$, $\mathbf u(t) \in \mathbb{R}^m$ and $\mathbf y(t) \in \mathbb{R}^p$. 
The function $\mathbf f: \mathbb{R}^n \rightarrow \mathbb{R}^n$ represents a Lipschitz continuous system function, 
$\mathbf h: \mathbb{R}^n \rightarrow \mathbb{R}^p$ a Lipschitz continuous output function, 
and $\mathbf g(\mathbf x(t)) \in \mathbb{R}^{n \times m}$ is referred to as the control effectiveness matrix. 
In the field of flight control, inversion-based control methodologies, such as NDI, are commonly used to impose desired closed-loop behavior on the MIMO dynamics \cite{Holzapfel_2004}. 
To do so, the time derivatives of the introduced nonlinear dynamics from Eq.~\eqref{NonlinearSystem_1} and Eq.~\eqref{NonlinearSystem_2} are derived and reformulated using the Lie derivative notation \cite{Lombaerts_2012}:
\begin{equation}
\label{eqn: time derivation of nonlinear control affine system}
\begin{split}
\dot{\mathbf y}(t) &= \frac{\partial \mathbf h}{\partial \mathbf x} \dot{\mathbf x}(t) 
= \frac{\partial \mathbf h}{\partial \mathbf x} \left[\mathbf f(\mathbf x(t))+\mathbf g(\mathbf x(t))\mathbf u(t)\right] \\
&= \frac{\partial \mathbf h}{\partial \mathbf x}\,\mathbf f(\mathbf x(t)) 
+ \frac{\partial \mathbf h}{\partial \mathbf x}\,\mathbf g(\mathbf x(t))\,\mathbf u(t) \\
&= L_{\mathbf f} \mathbf h(\mathbf x(t)) + L_{\mathbf g} \mathbf h(\mathbf x(t))\,\mathbf u(t) \\
&= \mathbf F(\mathbf x(t)) + \mathbf G(\mathbf x(t))\,\mathbf u(t).
\end{split}
\end{equation}
A relationship between the input and the output can be established if $\mathbf G(\mathbf x(t)) \neq 0$. 
For cases in which $\mathbf G(\mathbf x(t))=0$, the relationship must be further derived until the system input affects the system output. 
The number of time derivations needed is called the relative degree and is represented by $r$ \cite{Bhardwaj_2021}. 
For the sake of brevity, we consider here an example of a first-order system. 
For such a system with a relative degree of one, the NDI-based control law can be stated as: 
\begin{equation}
\mathbf u(t) = \mathbf G(\mathbf x(t))^{-1} \big[\bm\nu(t) - \mathbf F(\mathbf x(t))\big],
\end{equation}
with $\bm\nu(t) \in \mathbb{R}^o$ being a virtual control input vector created by a linear controller to impose the desired closed-loop characteristics on the controlled variables such that $\bm\nu(t) = -\mathbf K \mathbf y(t)$, with $\mathbf K$ being a Hurwitz gain matrix. 
It is evident that the inversion of the control effectiveness matrix $\mathbf G(\mathbf x(t))$, from now on for simplicity referred to as $\mathbf B$, maps a desired $\bm\nu(t)$ into the control space $\mathbb{R}^m$. 
In cases in which $m = n$ and $\operatorname{rank}(\mathbf B) = n$, this leads to a unique solution. 
In the case of over-actuation, where $m > n$, the inversion of the matrix is not easily available, since the mapping into the control space is not unique and, mathematically speaking, under-determined. 
In the flight control literature this problem is often referred to as the control allocation problem \cite{Petersen_2005}. 
In general, the control allocation problem can be interpreted as a constrained optimization problem since magnitude and rate limitations on each actual available control effector $u_i(t)$ need to be considered. 
The rate limits are often defined as symmetric lower and upper bounds for each actuator, leading to the following rate saturation function for each actuator $i$:
\begin{equation}
\label{rate_limits}
R(\dot{u}_i(t)) = \begin{dcases*}
\dot{u}_{\max,i},   & if $ \dot{u}_i(t)  > \dot{u}_{\max,i}$, \\
\dot{u}_{\min,i},   & if  $ \dot{u}_i(t) < \dot{u}_{\min,i}$,\\
\dot{u}_i(t),       & else.
\end{dcases*}
\end{equation}
Since a small and constant time step $T$ for the control computation can be assumed, the rate limits can be interpreted as additional rate-dependent magnitude limits $\overline{u}_{i}(t)$ relative to the current deflection state $u_i(t)$, modeled as a zero-order hold limit:
\begin{equation}
\overline{u}_{i}(t) = u_i(t-T) + R(\dot{u}_i(t))\, T.
\label{zero-order hold magnitude limit}
\end{equation}
Based on that, the magnitude limitations can be defined as individual lower and upper bounds for each effector $i$.  
The magnitude-constrained control inputs are represented using the following magnitude saturation function:
\begin{equation}
\label{magnitude_limits}
S(u_i(t),\dot{u}_i(t)) = \begin{dcases*}
\min(u_{\max,i},\overline{u}_{i}(t)),   & if  $ u_i(t)  \geq \min(u_{\max,i},\overline{u}_{i}(t))$, \\
\max(u_{\min,i},\overline{u}_{i}(t)),   & if  $ u_i(t) < \max(u_{\min,i},\overline{u}_{i}(t))$,\\
u_i(t),   & else.
\end{dcases*}
\end{equation}

\begin{figure}[ht]
\centering
\includegraphics[width=0.7\columnwidth]{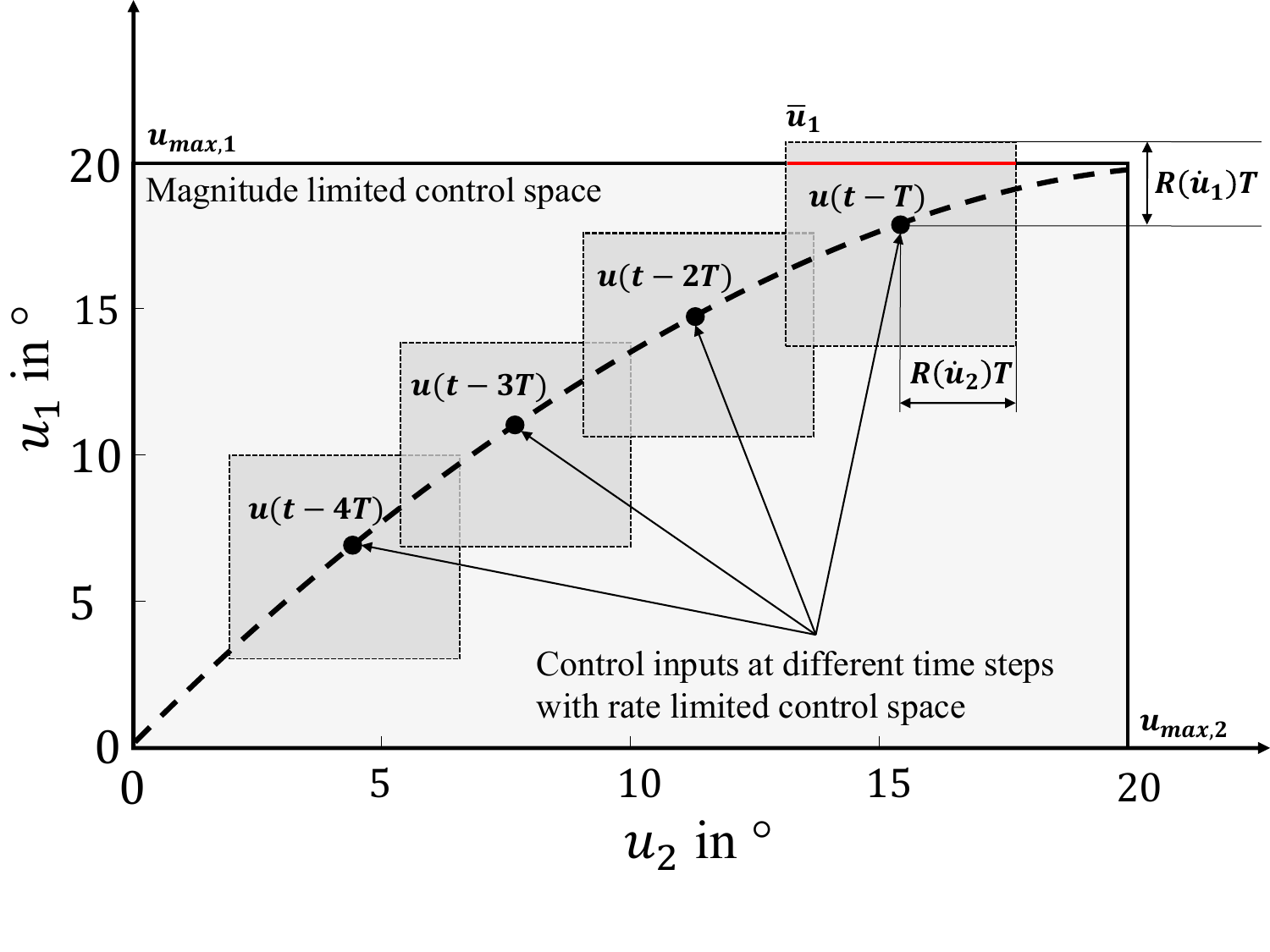}
\caption{Illustration of the discussed control allocation problem with considered magnitude and rate limits for a system with two control inputs $u_1$ and $u_2$.}
\label{fig:Control_space_constraints}
\end{figure}

This leads to the following formulation of the feasible set of control inputs $\boldsymbol{U}$:
\begin{equation}
\label{control_input_limits}
\boldsymbol{U} := \{\mathbf u(t) \in \mathbb{R}^{m} \;|\; \forall i \in [1,m] : \max(u_{\min,i},\overline{u}_{i}(t)) \leq u_i(t) \leq \min(u_{\max,i},\overline{u}_{i}(t))\}.
\end{equation}

The described admissible control input space for a static toy example with two magnitude- and rate-constrained control inputs $u_1$ and $u_2$ is given in Figure~\ref{fig:Control_space_constraints}. 

Based on the feasible set of control inputs, a feasible physical control volume $\boldsymbol D$ can be constructed as
\begin{equation}
\boldsymbol D := \{ \bm\nu(t) \in \mathbb{R}^o \;|\; \bm\nu(t) = \mathbf B \mathbf u(t), \;\mathbf u(t) \in \boldsymbol U \}.
\end{equation}

\begin{figure}[ht]
\centering
\includegraphics[width=0.5\columnwidth]{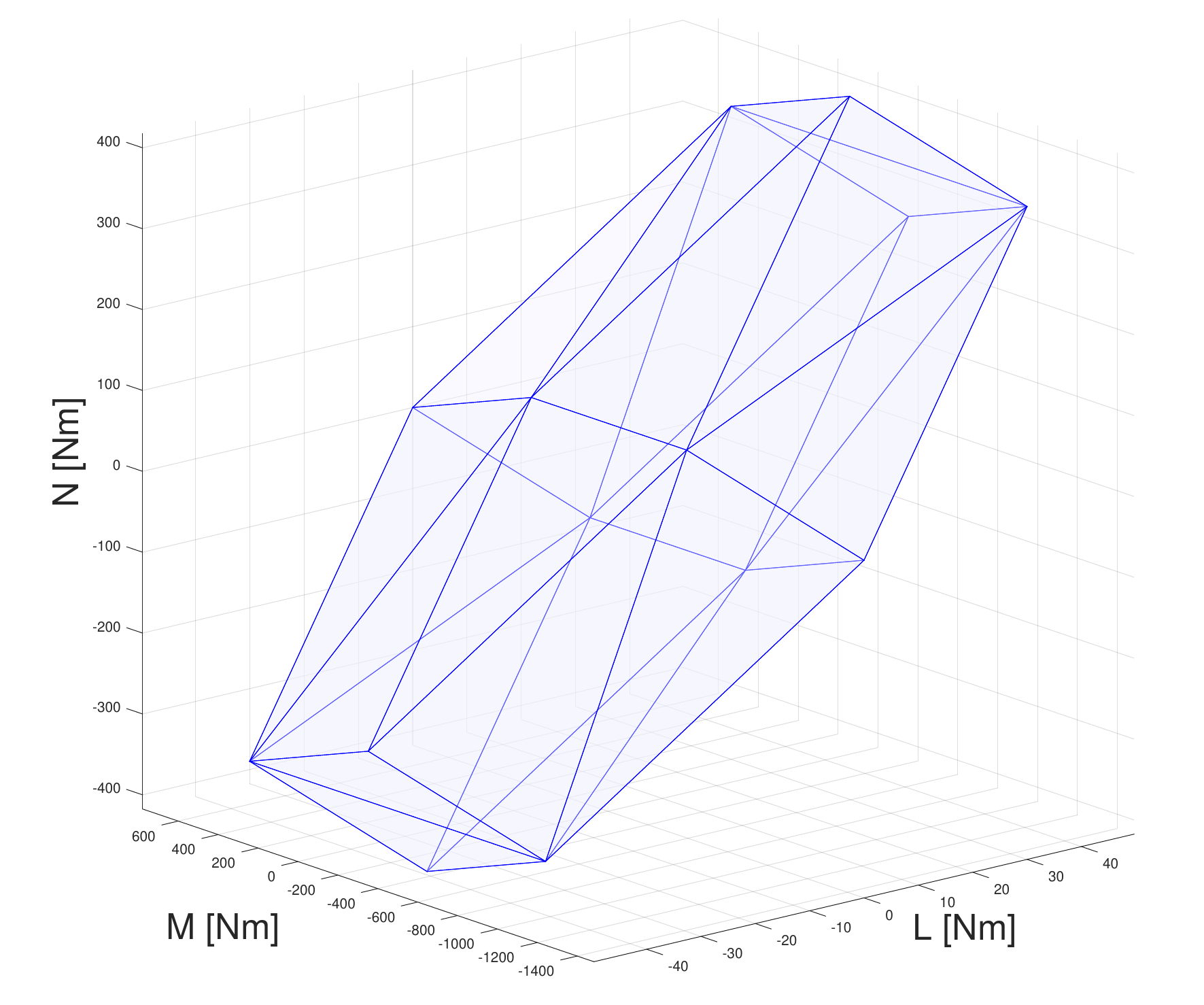}
\caption{An exemplary plot of an attainable moment set of the GHGV-2 for an operating point at $M = 8$ and $H= 30\,\text{km}$.}
\label{fig:AttainableMomentSet_GHGV2}
\end{figure}

In the context of flight control, the resulting convex set $\boldsymbol D$ is often referred to as the attainable moment set (AMS) \cite{Pfeile2021}. 
An example of an AMS, in which dynamic effects coming from the rate limits are neglected, for a hypersonic glide vehicle with four control effectors ($m=4$) is shown in Figure~\ref{fig:AttainableMomentSet_GHGV2}.

The resulting general control allocation problem, considering magnitude and rate limits on the control input, can be formulated as the following constrained optimization problem:
\begin{alignat}{2}
&\!\argmin_{\mathbf u(t) \in \mathbb{R}^{m}}      &\qquad&  \lVert  \mathbf u(t)\rVert_{p}\label{General_CA_problem}\\
&\text{subject to} &      & \bm\nu(t) = \mathbf B \mathbf u(t), \notag\\
&                  &      & \max(\mathbf u_{\min},\overline{\mathbf u}(t)) \leq \mathbf u(t) \leq \min(\mathbf u_{\max},\overline{\mathbf u}(t)), \notag
\end{alignat}
{\color{black}
with $p$ being a real number used to define the considered norm. The choice of the norm has a direct influence on the resulting control effort distribution and how the available actuator suite is used. In later sections of this work, the $\ell_2$-norm is adopted to promote a balanced distribution of control effort for the GHGV-2 and to enable an efficient least-squares formulation of the control allocation problem. Alternative formulations based on the $\ell_1$-norm or the $\ell_\infty$-norm are in many application cases equally well suited to handle actuator limitations, coupling effects, and additional preference criteria, and have been successfully applied and tested in similar flight control applications \cite{Schierman_2004}. Nevertheless, in order to provide a more comprehensive context for the choice of the $\ell_2$-norm adopted in this work, the influence of different norm selections on the solution of the control allocation problem is discussed in the following subsection.
}
{\color{black}
\subsection{The role of norms in control allocation}
\label{The role of norms in control allocation}
By interpreting the optimization problem from a geometric perspective, different norms induce different level sets in the control input space, which interact with the admissible control input set $\boldsymbol U$ and the AMS $\boldsymbol D$ in distinct ways. As a result, the choice of norm implicitly encodes a preference for certain control distribution strategies and affects how a demanded virtual control input $\bm{\nu}(t)$ is distributed among the available actuators.

The minimization of the Euclidean $\ell_2$-norm corresponds to a least-squares formulation and leads to a quadratic programming (QP) control allocation problem. The $\ell_2$-norm promotes a smooth and balanced distribution of control effort across all available actuators and yields a unique solution whenever the feasible set is non-empty \cite{harkegaard2003}. Due to its favorable numerical properties and balanced way of distributing the control effort within the available actuators, $\ell_2$-based control allocation has become a standard choice in many flight control applications \cite{Johansen_2004,Ola2004}. In contrast, the $\ell_1$-norm promotes sparsity in the control input vector and tends to concentrate control effort on a reduced subset of actuators. This property can be advantageous in systems where certain effectors are preferred due to efficiency or authority. Control allocation problems based on the $\ell_1$-norm can be formulated as linear programs and efficiently solved using simplex-based or interior-point methods. Linear programming (LP) control allocation approaches have been successfully demonstrated in flight-tested applications, including re-entry vehicles with actuator constraints \cite{Schierman_2004,Ikeda_2000, Bolender_2004}. The $\ell_\infty$-norm minimizes the maximum absolute actuator deflection within the actuator suite and can be interpreted as a worst-case input allocation strategy. This formulation is particularly suitable when limiting peak actuator usage is critical. Similar to the $\ell_1$-norm, $\ell_\infty$-based control allocation problems can be cast as linear programs \cite{Petersen_2005,Bodson_2011,Huang_2019}. 

It is worth noting that, for low-dimensional allocation problems or operating regimes in which actuator constraints remain inactive, some of the listed norm choices may yield very similar or in certain cases identical solutions. Pronounced differences typically arise in strongly constrained or highly over-actuated systems, where the geometry of the feasible control set interacts nontrivially with the selected norm. 

\begin{figure}[ht]
\centering
\includegraphics[width=\columnwidth]{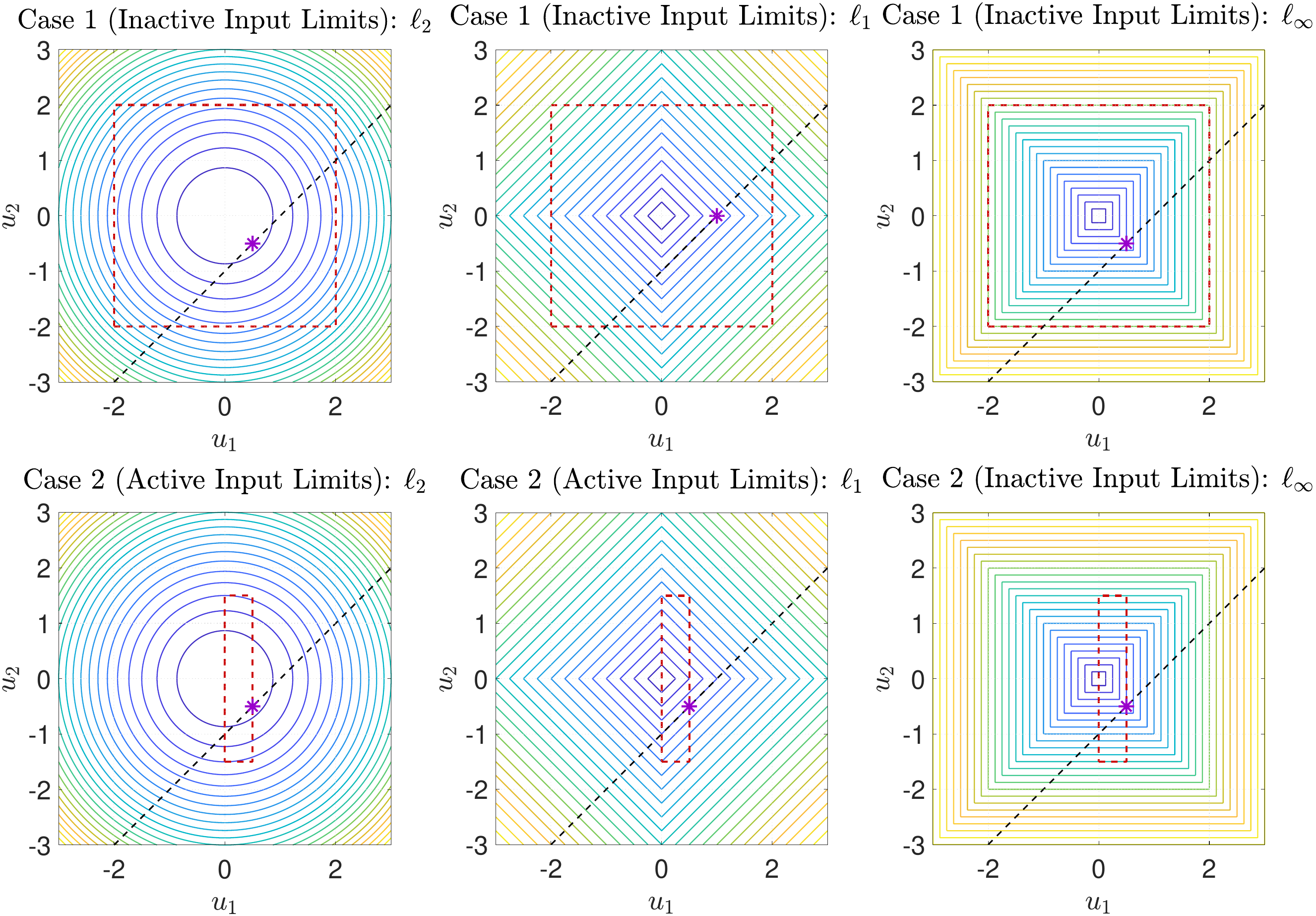}
\caption{{\color{black}Results of a comparative analysis illustrating the influence of different norm choices on control allocation with inactive and active input constraints.}}
\label{fig:Norm Comarison example control allocation}
\end{figure}

To further illustrate the influence of the chosen norm on the solution of constrained control allocation problems, a simplified static example is considered and visualized in Figure~\ref{fig:Norm Comarison example control allocation}. The problem involves a system with a two–dimensional control input vector $\mathbf{u} = [u_1,\,u_2]^T$, whose admissible set is bounded by magnitude constraints $\mathbf{u}_{\min} \leq \mathbf{u} \leq \mathbf{u}_{\max}$ and indicated by the dashed red box in the figure. The control effectiveness matrix is defined as $\mathbf{B} = [1,\,-1]$, and the demanded virtual control input is set to $\nu = 1$. The set of all control inputs satisfying the virtual control demand $\nu = \mathbf{B}\mathbf{u}$ forms a straight line in the $(u_1,u_2)$-plane and is shown as the dashed black line in Figure~\ref{fig:Norm Comarison example control allocation}. Two representative cases are examined. In \emph{Case~1}, the magnitude bounds are chosen as $\mathbf{u}_{\min} = [-2,\,-2]^T$ and $\mathbf{u}_{\max} = [2,\,2]^T$, such that the admissible region fully contains the unconstrained minima of the respective cost functions for the commanded $\nu$ of the case. Consequently, the input constraints are inactive, and the solution is determined solely by the norm-induced objective. In \emph{Case~2}, the admissible region is deliberately reduced by selecting $\mathbf{u}_{\min} = [0,\,-1.5]^T$ and $\mathbf{u}_{\max} = [0.5,\,1.5]^T$, such that only a limited segment of the solution line remains feasible. In this case, the input constraints are active and directly influence the resulting allocation. For each case, three objective functions of the form \( J = \lVert \mathbf{u} \rVert_p \) are considered, where the choice of \( p \) corresponds to the \( \ell_2 \)-, \( \ell_1 \)-, or \( \ell_\infty \)-norm. The corresponding optimization problems are formulated using the appropriate numerical frameworks for each norm, as discussed in the previous part of this section. In all cases, the equality constraint $\nu = \mathbf{B}\mathbf{u}$ as well as the actuator magnitude limits are enforced as hard constraints within the optimization, as presented in Eq.~\eqref{General_CA_problem}. The resulting cost contours are visualized in Figure~\ref{fig:Norm Comarison example control allocation}, enabling a direct geometric interpretation of the norm-induced allocation behavior.

For the $\ell_2$-norm, the cost contours are circular, and the optimal solution corresponds to the feasible point on the solution line that minimizes the Euclidean distance to the origin. In \emph{Case~1}, this yields the solution $\mathbf{u} = [0.5,\,-0.5]^T$, resulting in a balanced distribution of control effort between the two inputs. In \emph{Case~2}, the unconstrained minimum lies on the boundary of the admissible set, and the QP-based implementation is able to identify this solution, again resulting in $\mathbf{u} = [0.5,\,-0.5]^T$. In contrast, the $\ell_1$-norm yields diamond-shaped cost contours, corresponding to squares that are rotated by $45^\circ$ with respect to the coordinate axes in the $(u_1,u_2)$-plane. As a consequence, the associated optimization problem exhibits a sparsity-promoting behavior, tending to concentrate the control effort on a single actuator whenever this is compatible with the constraints. In \emph{Case~1}, the $\ell_1$-based allocation yields the solution $\mathbf{u} = [1,\,0]^T$, where the entire control effort is assigned to $u_1$ while $u_2$ remains inactive. This behavior highlights the tendency of $\ell_1$-based approaches to reach the magnitude saturation of the most effective actuators at an earlier stage, which can be undesirable in applications where actuator saturation is problematic or where a balanced usage of available actuators is preferred. Although this effect can be reduced through appropriate conditioning and weighting of the optimization problem, as discussed in \cite{Schierman_2004}, the underlying numerical tendency remains inherent to the $\ell_1$-norm formulation. In \emph{Case~2}, the active input limits significantly restrict the admissible solution space, such that the unconstrained $\ell_1$-minimum becomes infeasible. Nevertheless, the LP-based allocation routine is able to identify the same feasible solution as obtained with the other norms, namely $\mathbf{u} = [0.5,\,-0.5]^T$. This illustrates that, as long as the solution space is nonempty, all considered optimization formulations are capable of identifying admissible solutions, even when the cost function alone does not yield a unique or well-conditioned minimum. The $\ell_\infty$-norm leads, similar to the $\ell_1$-norm, to square-shaped cost contours but aligned with the coordinate axes. For the considered example, the resulting solution is identical to the $\ell_2$-based solution in both scenarios, yielding $\mathbf{u} = [0.5,\,-0.5]^T$ in \emph{Case~1} and \emph{Case~2}. This behavior can be attributed to the well conditioned structure of the considered control effectiveness matrix, where both control inputs contribute with comparable magnitude to the virtual control demand. Under such conditions, minimizing the peak control effort does not introduce a strong bias toward a particular actuator, and the $\ell_\infty$- and $\ell_2$-based solutions coincide.

A notable distinction between the considered norm formulations arises from their smoothness properties. While the $\ell_2$-norm defines a smooth and continuously differentiable objective function, both the $\ell_1$- and $\ell_\infty$-norms are non-smooth and exhibit sharp edges in their level sets \cite{harkegaard2003}. As demonstrated by the simplified example, these geometric properties directly influence the resulting allocation behavior. In particular, the $\ell_1$-norm exhibits a pronounced tendency to drive highly effective actuators into saturation, even in cases where alternative feasible solutions with more balanced actuator usage exist. Although this sparsity-promoting behavior can be beneficial in certain applications, it is generally undesirable in flight control systems where early saturation of individual actuators may reduce robustness or limit available control authority. The $\ell_\infty$-norm, on the other hand, in some cases yields solutions similar to those obtained with the $\ell_2$-norm when the control effectiveness of each actuator is comparable, as observed in the presented example. However, due to its non-smooth cost structure similar to $\ell_1$, $\ell_\infty$-based formulations may still lead to non-smooth allocation behavior and abrupt changes in the optimal solution when the active set changes \cite{Frost_2009,Frost_2010}. In dynamic settings, such effects can result in rapid switching of actuator commands, input chattering and frequent operation near rate limits, which is undesirable from both a performance and energy conservation perspective. While the inclusion of explicit rate constraints and adjustments on the numerical optimization structure can mitigate these effects to some extent, the underlying tendency toward non-smooth allocation remains inherent to the formulation.

Based on these considerations, an $\ell_2$-norm-based formulation is adopted in this work for the GHGV-2 application. The $\ell_2$-norm promotes a balanced distribution of control effort in accordance with actuator effectiveness and exhibits favorable smoothness properties in the presence of actuator constraints. Building on this formulation, a variety of control allocation algorithms that leverage the $\ell_2$-norm have been proposed and applied in the literature. These approaches differ in how the control allocation problem is structured and simplified, and in how actuator limitations are addressed within the allocation process. The following subsection provides a brief overview of commonly used $\ell_2$-based control allocation algorithms and discusses their respective advantages and limitations in the context of hypersonic glide vehicle applications.

}

\subsection{Common $\ell_2$-based Control Allocation Algorithms}
\label{Control Allocation Algorithms}
This section presents a brief overview of three commonly used algorithms in control allocation, highlighting their benefits and limitations in distributing virtual control input commands for the considered class of overactuated hypersonic systems.

\subsubsection{Pseudoinverse-Based Control Allocation}

The majority of the control allocation problems are solved by pseudoinverse-based control allocation (PICA) using the Moore-Penrose pseudoinverse, which implicitly solves the following unconstrained least-square problem for the allocation of the incremental virtual control input vector $\bm\nu$:
\begin{alignat}{2}
&\!\min_{\mathbf u \in \mathbb{R}^{m}}      &\qquad& \lVert  \mathbf u\rVert_{2}\\
&\text{subject to} &      & \bm\nu = \mathbf B \mathbf u,\notag
\end{alignat}
with $\| \cdot \|_2$ being the $\ell_2$-norm. The defined optimization problem has the following closed-form solution:
\begin{equation}
\mathbf u =  \mathbf B^{+} \bm\nu = \mathbf B^{T}\left[\mathbf B \mathbf B^{T}\right]^{-1} \bm\nu.
\end{equation}
The major drawback of the generalized inverse approach is that the constraints on the control input $\mathbf u$ are neglected, leading to the computation of possibly infeasible control inputs.

\subsubsection{Redistributed Pseudoinverse-Based Control Allocation}

An alternative approach to solve the constrained control allocation problem was proposed by the authors in \cite{Virning_1994} and is called redistributed pseudoinverse-based control allocation (RPICA). 
The core idea of the methodology is to cope with the problem of limited control input authority by iterating over the AMS until a feasible control input command is found. 
The algorithm starts by generating a first solution for a control vector candidate using the pseudoinverse
\begin{equation}
\mathbf u^{(1)} = \mathbf B^+ \bm\nu^{(0)},
\label{eq:first_iteration_RPICA}
\end{equation}
with $\bm\nu^{(0)}$ being the initially desired commanded virtual control input vector from the controller. In the case where no elements of $\mathbf u^{(1)}$ are saturated and Eq.~\eqref{eq:first_iteration_RPICA} delivers the desired virtual control input vector $\nu^{(0)}$, $\mathbf u^{(1)}$ is the final solution. In cases where at least one element of $\mathbf u^{(1)}$ conflicts with the available control limitations and the desired virtual control input vector is not generated, the process continues. The achievable control input is defined as
\begin{equation}
\bm\nu^{(1)} = \mathbf B \, S(\mathbf u^{(1)}),
\end{equation}
with $S(\mathbf u^{(1)})$ being a simplified version of the magnitude saturation function defined in Eq.~\eqref{magnitude_limits}, since in most of the formulations in the existing literature the rate limits are neglected. 
The control effectiveness matrix $\mathbf B$ is then sparsified such that the matrix columns which correspond to the saturated actuators are zeroed. 
Also, the limits are adjusted to make sure that the final solution does not exceed the actual bounds:
\begin{equation}
\label{RPICA_limits_1}
\mathbf u_{\max}^{(j)} = \mathbf u_{\max}^{(j-1)} - \mathbf u^{(j)},
\end{equation}
\begin{equation}
\label{RPICA_limits_2}
\mathbf u_{\min}^{(j)} = \mathbf u_{\min}^{(j-1)} - \mathbf u^{(j)}.
\end{equation}
Now, the remaining pseudo-control vector is computed as
\begin{equation}
\bm\nu^{(j)} = \bm\nu^{(j-1)} - \mathbf B_s \, S(\mathbf u^{(j)}),
\end{equation}
with $\mathbf B_s$ being the sparsified control effectiveness matrix. 
In the next iteration step, the remaining pseudo-control vector is allocated using $\mathbf B_s$ and the adjusted limits as defined in Eq.~\eqref{RPICA_limits_1} and Eq.~\eqref{RPICA_limits_2}. 
That procedure is continued until either the desired pseudo-control vector has been achieved or the maximum number of iterations has been reached. 
In both cases, the final control vector is the sum of the increments of all iteration steps.

For MIMO systems, it is preferable to keep the directionality of the unconstrained input vector $\mathbf u^{(j)}$. 
To make that possible, the authors in \cite{Zhang_2018} proposed the redistributed scaling pseudoinverse control allocation (RSPICA) algorithm. 
The saturated control input vector is scaled within the method to ensure perfect direction alignment with the desired pseudo-control vector. 
The used scalar $a$ is chosen such that the element that exceeds its respective limit the most is saturated, while the others are not:
\begin{equation}
\mathbf u_{a}^{(j)} =  \underbrace{\min\left(1, \frac{l(u_1^{(j-1)})}{u_1^{(j)}}, \frac{l(u_2^{(j-1)})}{u_2^{(j)}}, \hdots, \frac{l(u_m^{(j-1)})}{u_m^{(j)}}\right)}_{a} \cdot \mathbf u^{(j)},
\end{equation}
with
\begin{equation*}
l(u_i^{(j-1)}) = 
\begin{dcases*}
u_{\max,i}^{(j-1)}, & if $u_i^{(j)} > 0$, \\
u_{\min,i}^{(j-1)}, & if $u_i^{(j)} < 0$. 
\end{dcases*}
\end{equation*}

Both RPICA and its scaled variant, RSPICA, are advantageous for implementation in embedded systems with limited computational resources due to their simplicity. 
RSPICA, in particular, was designed to maintain input vector directionality in MIMO systems under saturation conditions by scaling the requested vector with a factor $a$, thereby ensuring that control input constraints are respected. However, despite these advancements, both methods have notable limitations. 
Their reliance on the $\ell_2$-based least-squares approach, inherent in the Moore-Penrose pseudoinverse, can result in an inefficient use of the AMS, when the asymmetrical input constraints and strong inter-directional control authority of the control effectors are present. Additionally, for systems with asymmetric input limits where the lower input bound is zero, RSPICA may produce zero-scaled solutions, resulting in trivial control commands. These challenges are particularly problematic for the certain classes of hypersonic vehicle.

\subsubsection{Quadratic Programming-Based Control Allocation}
\label{Quadratic Programming-based Control Allocation}

The constrained control allocation problem can also be solved using a quadratic programming-based control allocation (QPCA) approach. 
Examples of that can be found in the work presented in \cite{Johansen_2004} and \cite{Ola_2004}. 
To discuss this, we introduce the general formulation of a quadratic programming problem
\begin{alignat}{2}
&\!\min_{\mathbf x}      &\qquad& \frac{1}{2} \mathbf x^T \mathbf P \mathbf x + \mathbf q^T \mathbf x\\
&\text{subject to} &      & \mathbf A \mathbf x \leq \mathbf b,\notag
\end{alignat}
with $\mathbf A \mathbf x \leq \mathbf b$ being a linear matrix inequality to constrain the solution of $\mathbf x$. 
In the context of control allocation, we often aim to solve the following constrained weighted least-squares problem:
\begin{alignat}{2}
&\!\min_{\mathbf u}      &\qquad& \frac{1}{2} \| \mathbf B \mathbf u - \bm\nu \|_{\mathbf W}^2\\
&\text{subject to} &      & \mathbf A_u \mathbf u \leq \mathbf b_u,\notag
\end{alignat}
with $\mathbf A_u \mathbf u \leq \mathbf b_u$ reflecting control limits, and $\| \mathbf y \|_{\mathbf W} = \sqrt{\mathbf y^T \mathbf W \mathbf y}$ being a weighted form of the Euclidean norm for a vector $\mathbf y$. 
We can introduce a conversion of the least-squares cost function to a quadratic programming objective by
\begin{equation}
\begin{split}
\| \mathbf B \mathbf u - \bm\nu \|_{\mathbf W}^2 &= (\mathbf B \mathbf u - \bm\nu)^T \mathbf W (\mathbf B \mathbf u - \bm\nu) \\
&= \mathbf u^T \mathbf B^T \mathbf W \mathbf B \mathbf u - 2 \mathbf u^T \mathbf B^T \mathbf W \bm\nu + \bm\nu^T \mathbf W \bm\nu.
\end{split}
\end{equation}
Since the term $\bm\nu^T \mathbf W \bm\nu$ is constant and independent of $\mathbf u$, it can be discarded. 
This yields
\begin{equation}
\| \mathbf B \mathbf u - \bm\nu \|_{\mathbf W}^2 \propto \mathbf u^T \mathbf B^T \mathbf W \mathbf B \mathbf u - 2 ( \mathbf B^T \mathbf W \bm\nu )^T \mathbf u.
\end{equation}
Multiplying by $\tfrac{1}{2}$ gives the final quadratic form
\begin{equation}
\frac{1}{2} \mathbf u^T \mathbf B^T \mathbf W \mathbf B \mathbf u - (\mathbf B^T \mathbf W \bm\nu)^T \mathbf u,
\end{equation}
which corresponds to a quadratic programming problem with
\begin{equation}
\mathbf P = \mathbf B^T \mathbf W \mathbf B, \qquad \mathbf q = - \mathbf B^T \mathbf W \bm\nu.
\end{equation}
The constrained QPCA problem can therefore be formulated as
\begin{alignat}{2}
&\!\argmin_{\mathbf u \in \mathbb{R}^{m}}      &\qquad& \frac{1}{2} \mathbf u^T \mathbf B^T \mathbf W \mathbf B \mathbf u - (\mathbf B^T \mathbf W \bm\nu)^T \mathbf u\\
&\text{subject to} &      & \mathbf A_u \mathbf u \leq \mathbf b_u,\notag
\end{alignat}
with $\mathbf A_u = [\mathbf I_{m \times m}, - \mathbf I_{m \times m}]^T$ and $\mathbf b_u = [\min(u_{\max,i},\overline{u}_{i}), -\max(u_{\min,i},\overline{u}_{i})]^T$. 
The formulated QPCA problem can be solved using a variety of methods, such as interior-point, active-set, or simplex \cite{Ola2004}. 
The advantage of the method is that it is able to provide solutions to complex control allocation problems. 
However, a significant drawback of the approach is the high computational complexity and with that the lack of real-time guarantees for certain embedded systems integrated in aerial vehicles, for example the system integrated into the GHGV-2.

%% file: Sections/Problem_Formulation.tex
The considered GHGV-2 is modeled as a rigid body, and the nonlinear flight dynamics of the HGV are based on classical Newtonian mechanics. 
\begin{figure}
\centering
\includegraphics[width=0.7\columnwidth]{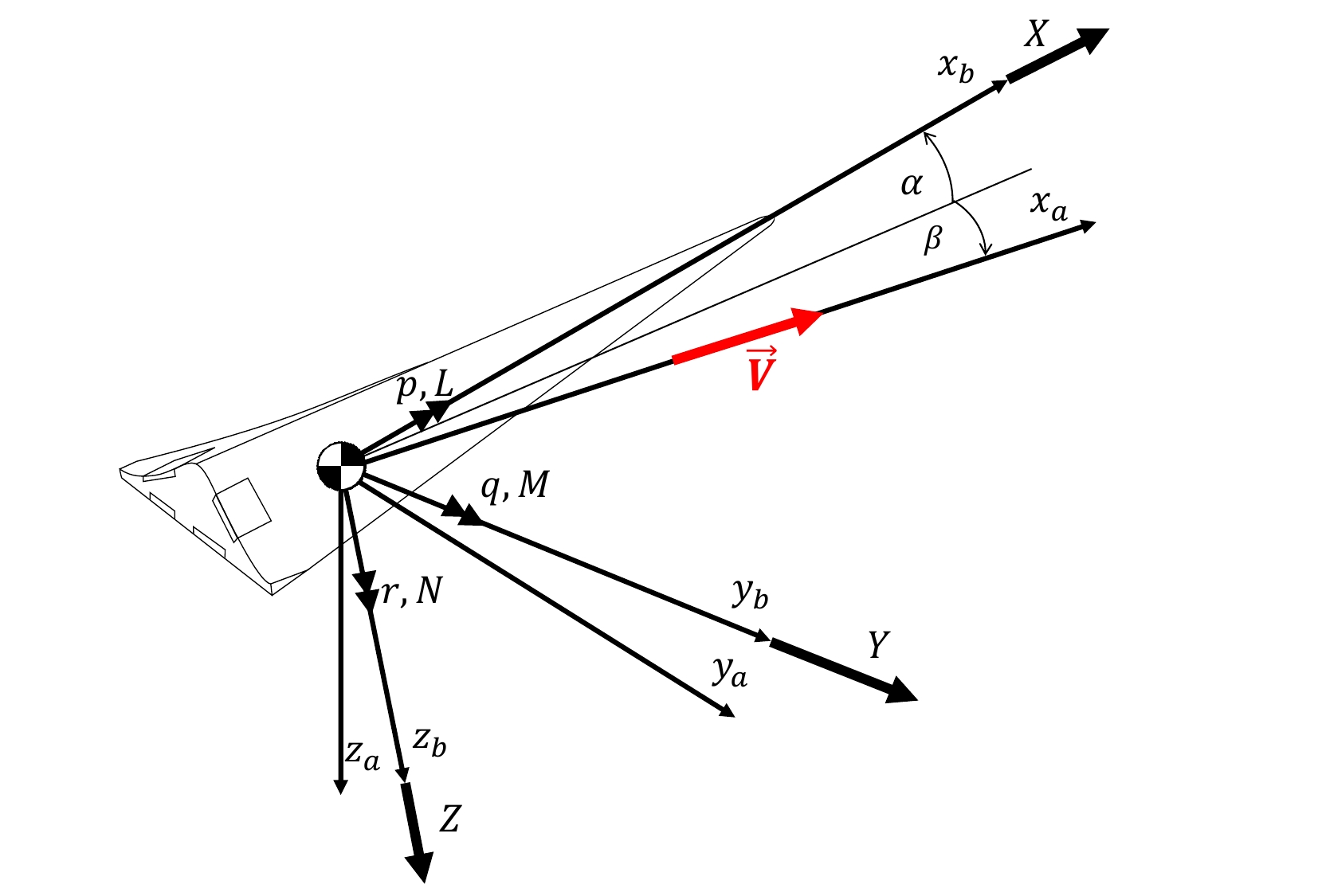}
\caption{Sketch of external forces and moments acting on the GHGV-2 concept.}
\label{fig:External_forces_and_moments}
\end{figure}
Figure~\ref{fig:External_forces_and_moments} displays the components \( X \), \( Y \), \( Z \) of the total external force vector and the components \( L \), \( M \), \( N \) of the total external moment vector expressed in the body-fixed frame of the vehicle. As detailed in Section~\ref{The General Control Allocation Problem}, the application of suitable control allocation algorithms is essential for the over-actuated GHGV-2, both in exoatmospheric and endoatmospheric flight phases. 

Due to the brevity of this paper, we limit our discussion to the control allocation problem during the endoatmospheric flight phase. In this context, only the aerodynamic and gravitational forces and moment effects are relevant for the considered problem and the later presented results.
\begin{figure}
\centering
\includegraphics[width=0.6\columnwidth]{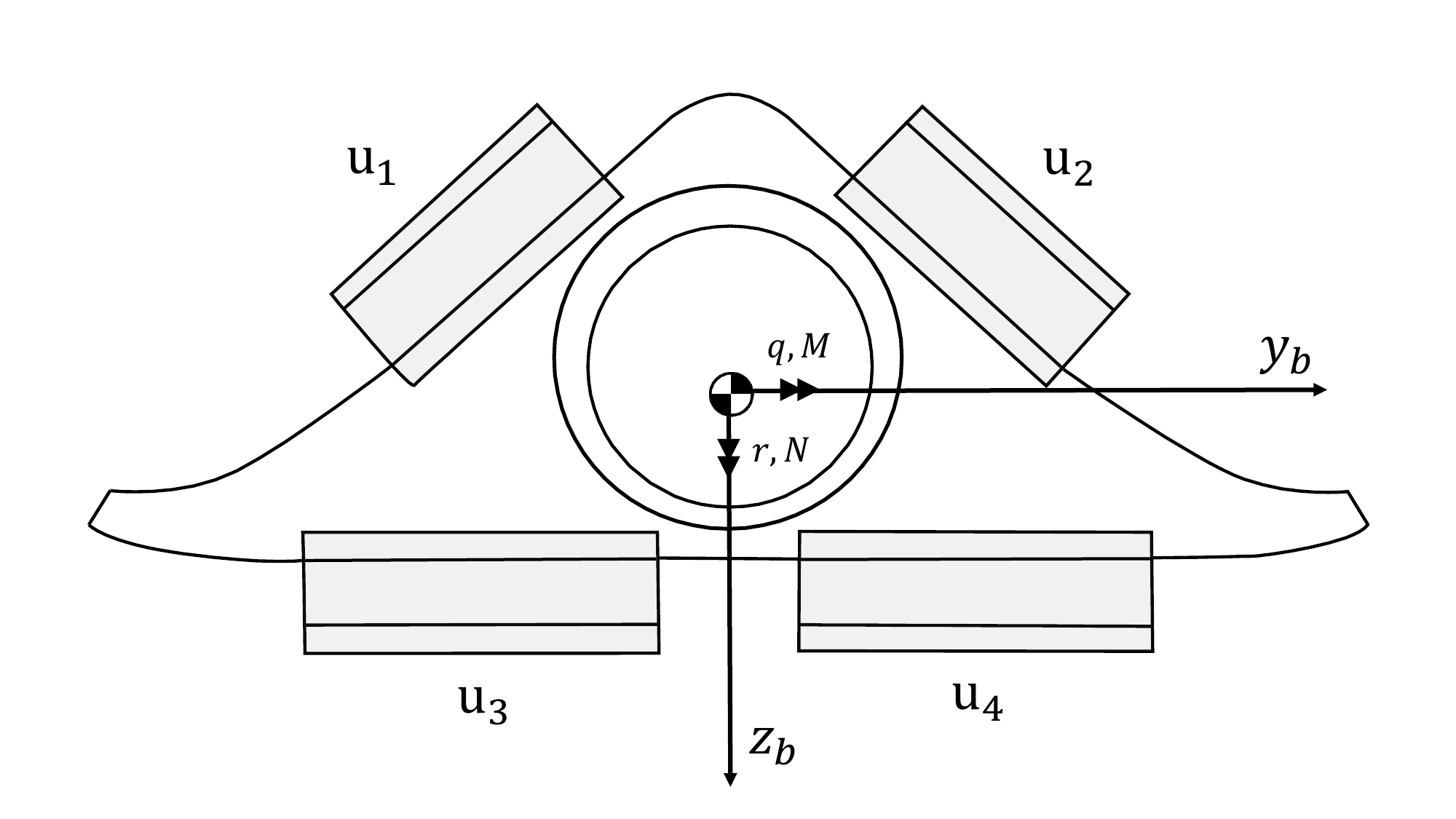}
\caption{Rear view of the GHGV-2 showing available control effectors during endoatmospheric operations.}
\label{fig:Available_control_effectors}
\end{figure}
Figure~\ref{fig:Available_control_effectors} illustrates the available aerodynamic control surfaces and their corresponding deflections: upper left fin \( u_{1} \), upper right fin \( u_{2} \), lower left fin \( u_{3} \), and lower right fin \( u_{4} \) ~\cite{Autenrieb2024}. This leads to the following control input vector:
\begin{equation}
\label{eqn:Delta_Vector}
\boldsymbol{u}(t) = \begin{bmatrix} u_{1}(t) & u_{2}(t) & u_{3}(t) & u_{4}(t) \end{bmatrix}^\mathrm{T}.
\end{equation}

The aerodynamic coefficient derivatives with respect to the control surface deflections can be fairly well approximated by linear terms. To account for the dependencies on the current flight state \( \boldsymbol{x}(t) \) and the operating point \( \bm{\sigma}(t)\), the control effectiveness matrix \( \mathbf{B}(\boldsymbol{x}(t),\bm{\sigma}(t)) \) is continuously re-computed using an onboard plant model and sensor measurements. However, it is assumed to be constant within each time step. The control effectiveness matrix \( \mathbf{B}(\boldsymbol{x}(t),\bm{\sigma}(t)) \) with respect to a virtual control command input vector \( \boldsymbol{\nu}(t) = [\nu_x(t), \nu_y(t), \nu_z(t)]^\mathrm{T} \) is thus defined as:
\begin{equation}
\mathbf{B}(\boldsymbol{x}(t),\bm{\sigma}(t)) = \begin{bmatrix}
\displaystyle \frac{\partial \nu_x}{\partial u_{1}} & \displaystyle \frac{\partial \nu_x}{\partial u_{2}} & \displaystyle \frac{\partial \nu_x}{\partial u_{3}} & \displaystyle \frac{\partial \nu_x}{\partial u_{4}} \\[8pt]
\displaystyle \frac{\partial \nu_y}{\partial u_{1}} & \displaystyle \frac{\partial \nu_y}{\partial u_{2}} & \displaystyle \frac{\partial \nu_y}{\partial u_{3}} & \displaystyle \frac{\partial \nu_y}{\partial u_{4}} \\[8pt]
\displaystyle \frac{\partial \nu_z}{\partial u_{1}} & \displaystyle \frac{\partial \nu_z}{\partial u_{2}} & \displaystyle \frac{\partial \nu_z}{\partial u_{3}} & \displaystyle \frac{\partial \nu_z}{\partial u_{4}}
\end{bmatrix},
\end{equation}
with \( \mathbf{B}(\boldsymbol{x}(t),\bm{\sigma}(t)) \in \mathbb{R}^{3 \times 4} \). As it is a non-square matrix, it cannot be inverted directly. A major challenge in hypersonic flights, often neglected in the literature, is the dependence of actuator magnitude and rate limits not only on the instantaneous flight state \( \boldsymbol{x}(t) \) but also on the operating point \( \bm{\sigma}(t) \). During high-speed atmospheric operations, high dynamic pressure induces counterforces on actuators, limiting the achievable deflections and rates. Nonlinear aerodynamic effects further complicate matters, with asymmetric influences on upper and lower surfaces requiring distinct actuator constraints. Additionally, operating-point-dependent thermal loads must be considered to minimize heating, reduce the burden on the TPS, and lower the infrared signature of the vehicle during operations.
\begin{figure}
\centering
\includegraphics[width=0.75\columnwidth]{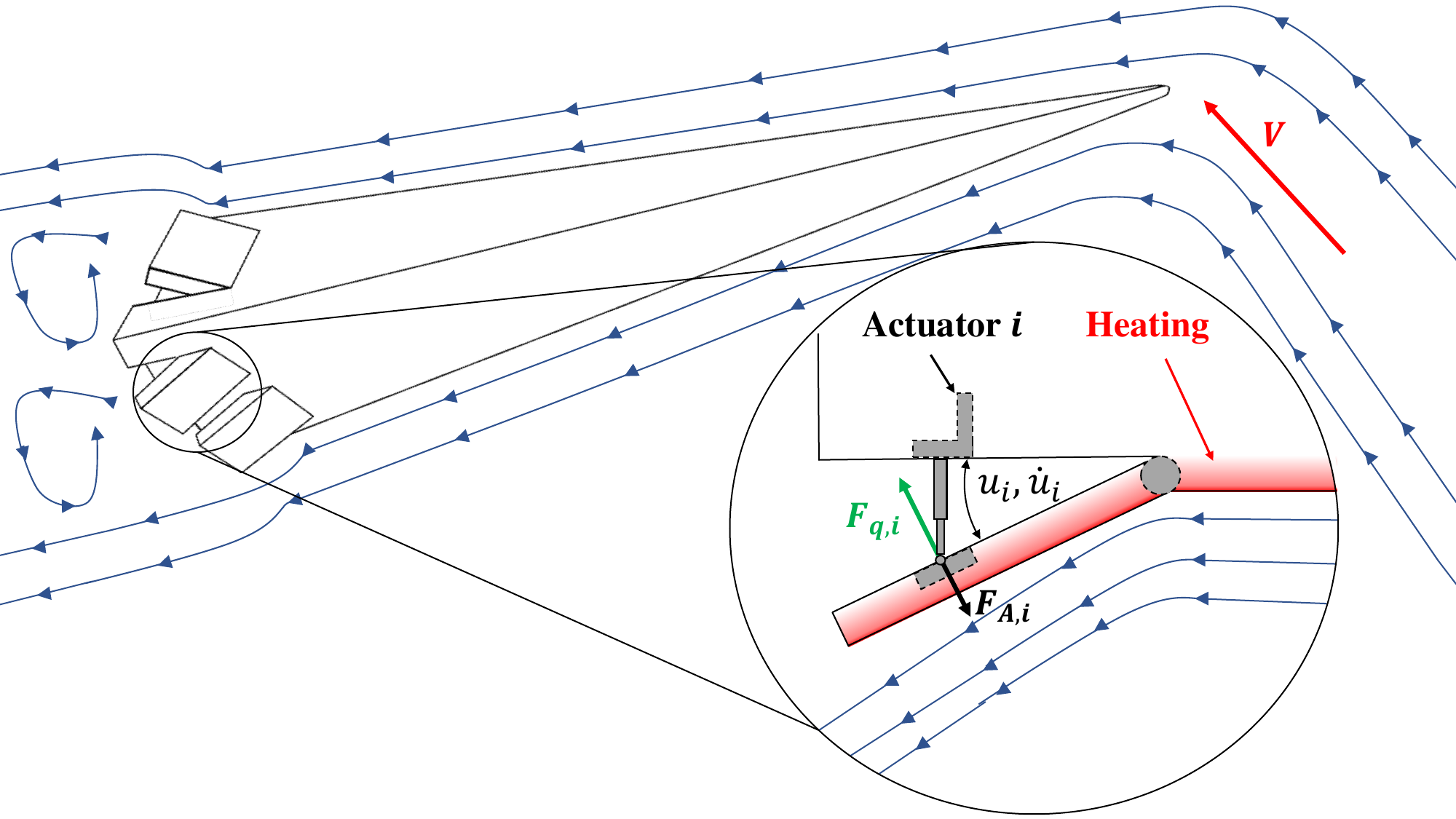}
\caption{Illustration of an overactuated hypersonic glide vehicle during high-speed atmospheric operations with high dynamic pressure and thermal loads.}
\label{fig:HGV_aerodynamic_thermal_effects}
\end{figure}
These effects are illustrated in Figure~\ref{fig:HGV_aerodynamic_thermal_effects}. In this figure, the red arrow \( \mathbf V \) represents the resulting velocity vector of the vehicle as it moves at high speed through the atmosphere. The blue arrows depict the airflow vector fields around the vehicle, shaping the aerodynamic forces and heating effects on the control surfaces. The control surface \( i \), modeled as a flapping plate, is characterized by a deflection angle \( u_i \) and a deflection rate \( \dot{u}_i \). The red gradient along the control surface indicates aerodynamic heating, generated by the vehicle's hypersonic speed, causing significant heat flux \( \tau_i \) on the surface. These heating effects must be considered during operation to avoid material degradation and structural failure of the control surfaces and to minimize the infrared signature in defense scenarios. The green arrow \( F_{q, i} \) represents the normal of the aerodynamic force acting on the surface, which is highly dependent on the deflection angle and the dynamic pressure \( q = 0.5 \rho V^2 \) acting on the control surface area \( S_{u, i} \), with \( \rho \) being the atmospheric density and \( V \) the resulting velocity. The actuator force \( \mathbf F_{A, i} \) is responsible for controlling the position of the surface, working against \( \mathbf F_{q, i} \), especially under high dynamic pressure conditions. Under these conditions, the actuator's ability to adjust the deflection angle and rate is sometimes restricted, leading to cases in which the deflection angle \( u_i \) is limited such that the resulting dynamic pressure force \( F_{q, i} \) exceeds the maximum force that the actuator can provide.

To address these high Mach number effects, we define the control allocation problem as a constrained optimization problem with state- and operating-point-dependent magnitude and rate limitations. The rate saturation function for each actuator $i$ is defined as:
\begin{equation}
\label{rate_limits2}
R(\dot{u}_i(t), \boldsymbol{x}(t), \bm{\sigma}(t)) =
\begin{cases}
\dot{u}_{\max,i}(\boldsymbol{x}(t),\bm{\sigma}(t)), & \text{if } \dot{u}_i(t) > \dot{u}_{\max,i}(\boldsymbol{x}(t),\bm{\sigma}(t)), \\
\dot{u}_{\min,i}(\boldsymbol{x}(t),\bm{\sigma}(t)), & \text{if } \dot{u}_i(t) < \dot{u}_{\min,i}(\boldsymbol{x}(t),\bm{\sigma}(t)), \\
\dot{u}_i(t), & \text{else}.
\end{cases}
\end{equation}
By modeling a rate-dependent magnitude limit $\overline{u}_{i}$ relative to the current deflection state \( \boldsymbol{u}(t) \), via Eq.~\eqref{zero-order hold magnitude limit}, we can define a rate- and state-/operating-point-dependent magnitude saturation function:
\begin{equation}
\label{magnitude_limits2}
S(u_i(t), \dot{u}_i(t), \boldsymbol{x}(t), \bm{\sigma}(t)) =
\begin{dcases*}
\operatorname{min}(u_{\max,i}(\boldsymbol{x}(t),\bm{\sigma}(t)),\overline{u}_{i}(t)), & if $u_i(t) \geq \operatorname{min}(u_{\max,i}(\boldsymbol{x}(t),\bm{\sigma}(t)),\overline{u}_{i}(t))$, \\
\operatorname{max}(u_{\min,i}(\boldsymbol{x}(t),\bm{\sigma}(t)),\overline{u}_{i}(t)), & if $u_i(t) < \operatorname{max}(u_{\min,i}(\boldsymbol{x}(t),\bm{\sigma}(t)),\overline{u}_{i}(t))$, \\
u_i(t), & else.
\end{dcases*}
\end{equation}
Leading to the following definition of the state- and operating-point-dependent feasible set of control inputs \( \boldsymbol{U}(\boldsymbol{x}(t),\bm{\sigma}(t)) \):
\begin{equation}
\label{control_input_limits2}
\boldsymbol{U}(\boldsymbol{x}(t),\bm{\sigma}(t)) := \{\boldsymbol{u}(t) \in \mathbb{R}^{m} \;|\; \forall i \in [1,m] : \operatorname{max}(u_{\min,i}(\boldsymbol{x}(t),\bm{\sigma}(t)),\overline{u}_{i}(t)) \leq u_i(t) \leq \operatorname{min}(u_{\max,i}(\boldsymbol{x}(t),\bm{\sigma}(t)),\overline{u}_{i}(t))\}.
\end{equation}
{\color{black}

A control allocation problem is formulated that not only considers the current control distribution at each time step, but also includes the previous time step as a soft constraint. This leads to the following constrained state- and operating-point-dependent optimization problem:
\begin{alignat}{2}
&\!\min_{\boldsymbol{u}(t) \in \mathbb{R}^{m}}      &\qquad& \|\mathbf{W}_m (\boldsymbol{u}(t) - \boldsymbol{u}_s(t)) \|_2^2 + \|\mathbf{W}_r (\boldsymbol{u}(t) - \boldsymbol{u}(t-T)) \|_2^2\label{State_dependent_General_CA_problem}\\
&\text{subject to} &      & \boldsymbol{\nu}(t) = \mathbf{B}(\boldsymbol{x}(t),\bm{\sigma}(t))\,\boldsymbol{u}(t), \notag\\
&                  &      & \boldsymbol{u}(t) \in \boldsymbol{U}(\boldsymbol{x}(t),\bm{\sigma}(t)), \notag
\end{alignat}
with $\boldsymbol{u}_s(t)$ denoting a desired steady-state control input for a given virtual control command $\boldsymbol{\nu}(t)$. The vector $\boldsymbol{u}_s(t)$ represents a preferred steady-state solution of the algebraic allocation relation $\boldsymbol{\nu}(t)=\mathbf B(\mathbf x(t),\boldsymbol{\sigma}(t))\,\boldsymbol u(t)$ and serves as a reference point for the solution process. At this stage, $\boldsymbol{u}_s(t)$ is introduced as an auxiliary design quantity and is neither required to be unique nor enforced as a hard constraint. Its purpose is to encode designer preferences regarding steady-state actuator usage, for example with respect to load distribution or thermal considerations. The specific construction of $\boldsymbol{u}_s(t)$ and its integration within the subsequently proposed control allocation strategy are addressed in Section~\ref{Designer Preference-Based Steady State Solution}. The weighting matrix $\mathbf W_m$ penalizes deviations of the computed control input from the preferred steady-state solution, whereas $\mathbf W_r$ penalizes variations between consecutive control inputs $\boldsymbol u(t)$ and $\boldsymbol u(t-T)$, thereby promoting smooth actuator behavior. Design guidelines and problem-specific choices for the weighting matrices $\mathbf W_m$ and $\mathbf W_r$ for the control allocation problem of the GHGV-2 are provided in Section~\ref{Actuator State-Based Weighting Matrices}.

In summary, the problem addressed in this work is to determine, at each time instant, a control input $\boldsymbol{u}(t)$ for the system defined in Eqs.~\eqref{NonlinearSystem_1} and \eqref{NonlinearSystem_2}, in accordance with Eq.~\eqref{State_dependent_General_CA_problem}, such that the virtual control input demand
$\mathbf{B}(\boldsymbol{x}(t),\boldsymbol{\sigma}(t))\,\boldsymbol{u}(t)=\boldsymbol{\nu}(t)$
is satisfied whenever the command lies within the state- and operating-point-dependent actuator limits.
If the command is not feasible, the control input is determined so as to minimize the deviation between the requested and achievable virtual control input while respecting the input constraints defined in Eq.~\eqref{control_input_limits2}.
}

%% file: Sections/Control_Allocation_Concept.tex
As discussed in Section~\ref{Problem Formulation}, during atmospheric flight, HGVs face significant challenges in maintaining stability and controllability due to thermal loads and limited aerodynamic authority~\cite{Li_2017}. These constraints complicate the design of effective control strategies and require careful management of aerodynamic and thermal effects. Owing to this complexity, the nominal atmospheric flight trajectory of the HGV is typically computed offline prior to the mission. During this offline optimization process, an optimal control problem is solved to obtain a feasible reference trajectory that satisfies prescribed initial and terminal conditions as well as physical constraints~\cite{An_2020}.
\begin{figure}
\centering
\includegraphics[width=0.7\columnwidth]{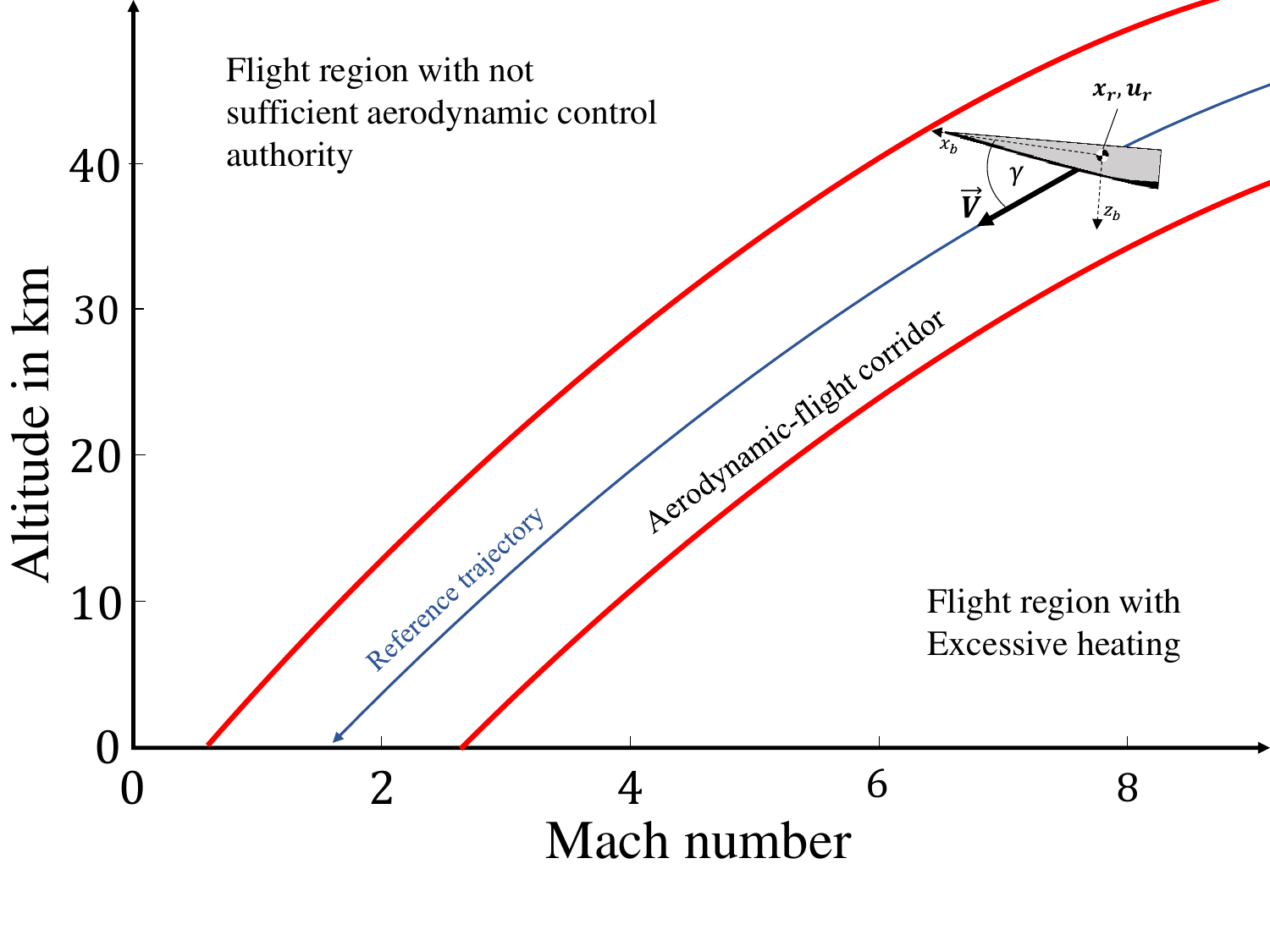}
\caption{Illustration of an operational flight envelope for a hypersonic glide vehicle, showing the reference trajectory bounded by upper and lower limits.}
\label{fig:Flight_Envelope}
\end{figure}
{\color{black}
The resulting optimized state reference trajectory is bounded by aerodynamic flight corridor constraints, as illustrated in Fig.~\ref{fig:Flight_Envelope} \cite{Li_2017}. Exceeding the upper altitude bound leads to insufficient aerodynamic control authority due to low air density, whereas descending below the lower bound exposes the vehicle to excessive aerodynamic heating, potentially causing material degradation or structural failure. Based on the obtained state reference trajectory $\boldsymbol{x}_r(t)$, an offline baseline control input $\boldsymbol{u}_r(t)$ is computed using a computationally demanding but accurate QPCA method, as discussed in Section~\ref{Quadratic Programming-based Control Allocation}, for the problem defined in Eq.~\eqref{State_dependent_General_CA_problem}. This offline solution incorporates all known hard and soft actuator constraints and is assumed to represent an optimal solution for a disturbance-free environment. The resulting control inputs are stored in lookup tables and serve as baseline commands during flight. Under nominal conditions, the baseline control input $\boldsymbol{u}_r(t)$ is sufficient to maintain the vehicle on its reference trajectory. However, due to model uncertainties, unmodeled dynamics, and external disturbances inherent to hypersonic flight, additional corrective control action is required. To address this issue, a nonlinear feedback controller generates a corrective virtual control input $\Delta \boldsymbol{\nu}(t)$ to compensate for deviations from the nominal trajectory, further information about the controller are provided in~\cite{Autenrieb2024}.

Assuming an affine structure of the control input, the total actuator command is expressed as
\begin{equation}
   \boldsymbol{u}(t) = \boldsymbol{u}_r(t) + \Delta \boldsymbol{u}(t).
\end{equation}
Here, $\Delta \boldsymbol{u}(t)$ represents an incremental control input required to realize the corrective virtual control command $\Delta \boldsymbol{\nu}(t)$ that is generated by the nonlinear feedback controller. Although the assumed affine nature of the control input formally allows the decomposition and superposition of the baseline command and the incremental correction input, the proposed iterative online control allocation algorithm does not allocate $\Delta \boldsymbol{u}(t)$ independently. Instead, it searches for the full control input $\boldsymbol{u}(t)$ while using the offline-optimized allocation solution $\boldsymbol{u}_r(t)$ as a baseline. This design choice leverages the fact that $\boldsymbol{u}_r(t)$ already represents a physically meaningful and near-feasible control input for the nominal operating case, thereby providing a well-conditioned starting point for the online allocation task. By incorporating $\boldsymbol{u}_r(t)$ directly into the allocation process, the search for a feasible control input is biased toward solutions that remain close to the nominal baseline, which reduces the likelihood of rapid input changes and converging to poorly conditioned allocations. This is particularly important in scenarios with strongly asymmetric actuator effectiveness or state-dependent magnitude and rate constraints, where the admissible solution space may be highly restricted. 
\begin{figure}
\centering
\includegraphics[width=0.8\columnwidth]{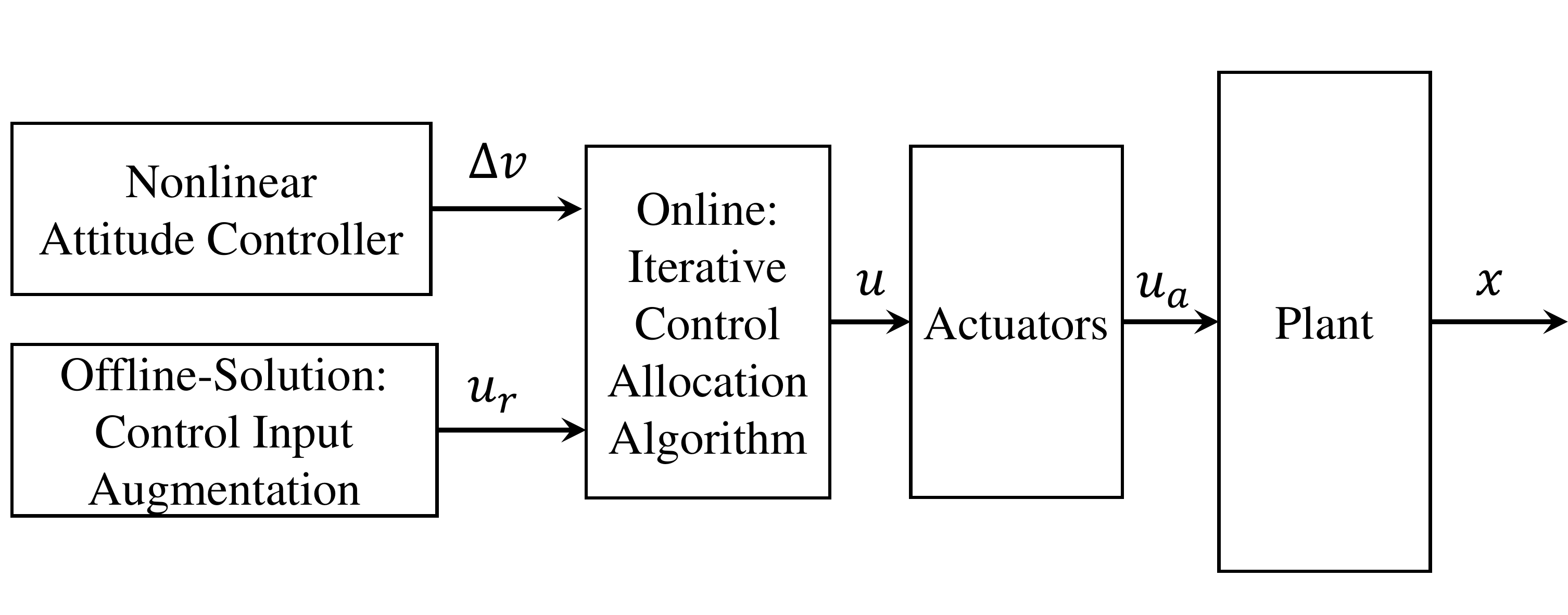}
\caption{Concept of the integrated control allocation system for the GHGV-2.}
\label{fig:Control_Allocation_Concept}
\end{figure}
The overall control allocation strategy is illustrated in Fig.~\ref{fig:Control_Allocation_Concept}. By combining an offline-optimized baseline solution with an online allocation mechanism operating on the full control input, the proposed architecture facilitates the real-time computation of feasible control commands in the presence of disturbances and uncertainties. The design and implementation of the iterative online control allocation algorithm, which constitutes a central contribution of this work, are detailed in the following section.

}

%% file: Sections/ProposedOnlineControlAllocationAlgorithm.tex
{\color{black}

In standard RPICA approaches, the problem of limited control input authority is addressed by iterating over the AMS until a feasible control input command is found \cite{Virning_1994}. During the iteration steps, the algorithm typically employs the Moore–Penrose pseudoinverse. The main advantage of this approach is that a solution can often be obtained in a simple and computationally efficient manner, making it suitable for real-time implementation in many aerospace applications. While it is, in principle, possible to account for actuator rate information indirectly through equivalent magnitude constraints, such considerations are typically neglected in the majority of the practical RPICA implementations. Moreover, most RPICA formulations do not adequately represent asymmetric input or rate limitations, particularly when these constraints are intended to act as soft shaping mechanisms that influence the control distribution prior to constraint activation. For the class of hypersonic glide vehicles studied here, these aspects are particularly of high interest, rendering standard RPICA solutions suboptimal.
}

In the proposed iterative dynamic control allocation (IDCA) algorithm, a linear filter formulation, originally introduced in \cite{Ola_2004}, is employed. Unlike the Moore--Penrose inverse, this approach explicitly accounts for both the current control distribution and the allocation of the previous time step as soft constraints \cite{Sadien_2020}. This implicitly incorporates actuator rates into the allocation problem. The method provides an explicit solution to the following weighted least-squares problem:
\begin{alignat}{2}
&\!\min_{\boldsymbol{u}(t) \in \mathbb{R}^{m}} &\qquad& \|\mathbf{W}_m(\boldsymbol{u}(t)-\boldsymbol{u}_s(t))\|_2^2 
+ \|\mathbf{W}_r(\boldsymbol{u}(t)-\boldsymbol{u}(t-T))\|_2^2 \label{dynamic_allocation_unsaturated}\\
&\text{subject to} &\qquad& \boldsymbol{\nu}(t) = \mathbf{B}\,\boldsymbol{u}(t), \notag
\end{alignat}
where $\boldsymbol{\nu}(t)=\mathbf{B}\,\boldsymbol{u}_r(t)+\Delta\boldsymbol{\nu}(t)$ is the virtual control input to be generated. The desired steady-state solution $\boldsymbol{u}_s(t)$ is specified in Section~\ref{Designer Preference-Based Steady State Solution}. The algorithm is designed so that, if $\boldsymbol{u}_s(t)$ is feasible, the steady-state solution converges to $\boldsymbol{u}_s(t)$ or to a solution close to it. The weighting matrices $\mathbf{W}_m$ and $\mathbf{W}_r$ are symmetric and defined as functions of the state and control; their detailed design is discussed in Section~\ref{Actuator State-Based Weighting Matrices}. Since $\mathbf{W}_m$ and $\mathbf{W}_r$ are symmetric, the combined weighting
\begin{equation}
    \mathbf{W} = \sqrt{\mathbf{W}_m^2 + \mathbf{W}_r^2}
\end{equation}
is non-singular. This allows the reformulation of Eq.~\eqref{dynamic_allocation_unsaturated} into the simpler criterion
\begin{alignat}{2}
&\!\min_{\boldsymbol{u}(t) \in \mathbb{R}^{m}} &\qquad& \|\mathbf{W}(\boldsymbol{u}(t)-\boldsymbol{u}_0(t))\|_2, \label{simpler_dynamic_allocation_unsaturated}
\end{alignat}
with
\begin{equation*}
    \boldsymbol{u}_0(t) = \mathbf{W}^{-2}\big(\mathbf{W}_m^2 \boldsymbol{u}_s(t) + \mathbf{W}_r^2 \boldsymbol{u}(t-T)\big).
\end{equation*}

The closed-form solution of Eq.~\eqref{dynamic_allocation_unsaturated}, expressed in the form of Eq.~\eqref{simpler_dynamic_allocation_unsaturated}, can be written as \cite{Ola_2004}:
\begin{equation}
\label{linear_filter_solution}
    \boldsymbol{u}(t) = \mathbf{E}\,\boldsymbol{u}_s(t) + \mathbf{F}\,\boldsymbol{u}(t-T) + \mathbf{G}\,\boldsymbol{\nu}(t),
\end{equation}
with
\begin{equation*}
    \mathbf{E} = (\mathbf{I}-\mathbf{G}\mathbf{B})\mathbf{W}^{-2}\mathbf{W}_m^2, \quad
    \mathbf{F} = (\mathbf{I}-\mathbf{G}\mathbf{B})\mathbf{W}^{-2}\mathbf{W}_r^2, \quad
    \mathbf{G} = \mathbf{W}^{-1}(\mathbf{B}\mathbf{W}^{-1})^{+}.
\end{equation*}

The explicit solution of Eq.~\eqref{linear_filter_solution} is used within an iterative procedure to enforce actuator constraints. The algorithm is executed within a fixed controller cycle, so external variables such as $\boldsymbol{u}(t-T)$ and $\boldsymbol{x}(t)$ are treated as constant during the inner iterations. A closely related line of work was presented by Burken et al.~\cite{Burken_2001}, who introduced a fixed-point, quadratic-programming-based iteration scheme for control allocation. Their method shares structural similarities with the iterative procedures considered here, in that both refine the solution by re-evaluating residual commands under actuator constraints. While the fixed-point scheme demonstrated promising performance for asymmetric re-entry vehicle configurations and provided a rigorous QP-based foundation for handling magnitude constraints, actuator rate limits were only mentioned as a possible extension and not explicitly included. By contrast, the algorithm presented in the following integrates rate-dependent limits directly into the allocation problem, both as explicit constraints and as soft penalties in the weighted least-squares cost. Moreover, the proposed formulation emphasizes practical implementation by illustrating how the weighting structure can incorporate operational considerations such as thermal load reduction and drag-sensitive allocation, and by providing empirical evidence in the simulation results that the method can robustly handle strongly time-varying, state- and operating-point-dependent actuator constraints.

\begin{figure}[ht!]
    \centering
    \includegraphics[width=0.7\textwidth]{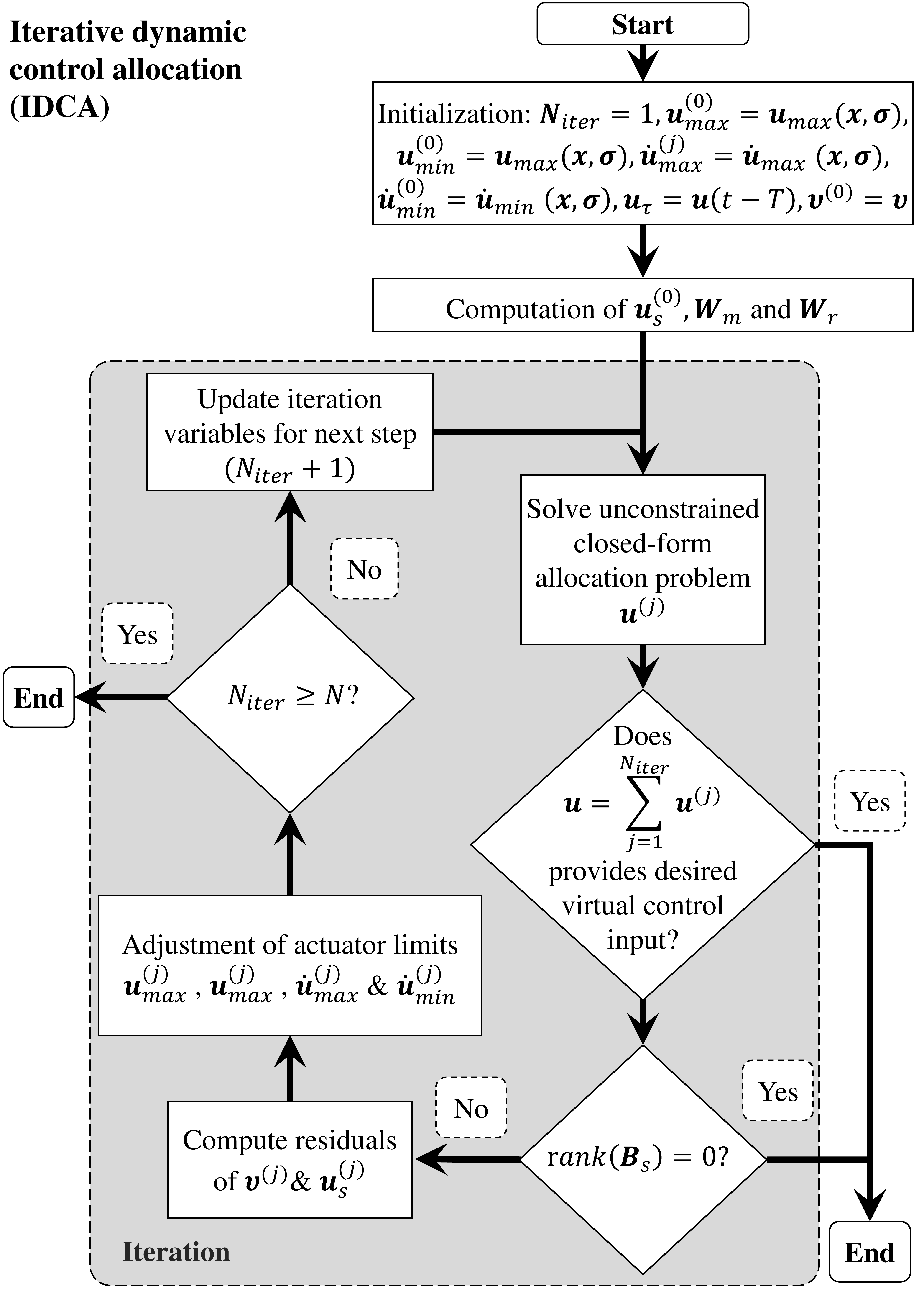}
    \caption{Flowchart of the proposed IDCA algorithm.}
    \label{fig:Control_Allocation_Process}
\end{figure}

Figure~\ref{fig:Control_Allocation_Process} illustrates the IDCA process. The iterative procedure allows a maximum number of $N$ iterations, with each step indexed by $(j)$. The actual number of iterations is denoted by $N_{\text{iter}} \leq N$. If a valid solution is found earlier, or if no further unsaturated actuators are available, the process terminates after $N_{\text{iter}}$ iterations.  

We define the iterative residual virtual control command as $\boldsymbol{\nu}^{(j)}$. In the first iteration step, the control input $\boldsymbol{u}^{(1)}$, corresponding to the initial residual $\boldsymbol{\nu}^{(0)}=\boldsymbol{\nu}(t)$, is computed using Eq.~\eqref{linear_filter_solution}. If $\boldsymbol{u}^{(1)}$ satisfies all actuator magnitude and rate constraints and realizes the desired virtual control input $\boldsymbol{\nu}(t)$, it is accepted as the final solution. Otherwise, the iterative redistribution procedure is continued. The achieved saturated virtual control in iteration step $j$ is
\begin{equation}
    \boldsymbol{\nu}^{(j)} = \mathbf{B}\,S(\boldsymbol{u}^{(j)},\dot{\boldsymbol{u}}^{(j)},\boldsymbol{x}(t)),
\end{equation}
where $S(\cdot)$ is the state- and rate-dependent saturation function (cf. Eq.~\eqref{magnitude_limits2}), $\dot{\boldsymbol{u}}^{(j)}=\tfrac{1}{T}(\boldsymbol{u}^{(j)}-\mathbf{u}_\tau)$ and $\mathbf{u}_\tau=\mathbf{u}(t-T)$ the control input of the last time step. If actuators saturate, the corresponding columns of $\mathbf{B}$ are set to zero, yielding the reduced matrix $\mathbf{B}_s$. Input limits are updated at each step to ensure feasibility:
\begin{align}
\boldsymbol{u}_{\max}^{(j)} &= \boldsymbol{u}_{\max}^{(j-1)} - \boldsymbol{u}^{(j)}, \qquad
\boldsymbol{u}_{\min}^{(j)} = \boldsymbol{u}_{\min}^{(j-1)} - \boldsymbol{u}^{(j)}, \\
\dot{\boldsymbol{u}}_{\max}^{(j)} &= \dot{\boldsymbol{u}}_{\max}^{(j-1)} - \tfrac{1}{T}(\boldsymbol{u}^{(j)}-\mathbf{u}_\tau), \qquad
\dot{\boldsymbol{u}}_{\min}^{(j)} = \dot{\boldsymbol{u}}_{\min}^{(j-1)} - \tfrac{1}{T}(\boldsymbol{u}^{(j)}-\mathbf{u}_\tau).
\end{align}

A corresponding adjustment of the steady-state solution $\boldsymbol{u}_s$ is also applied. The residual virtual control vector is then updated as
\begin{equation}
    \boldsymbol{\nu}^{(j)} = \boldsymbol{\nu}^{(j-1)} - \mathbf{B}_s S(\boldsymbol{u}^{(j)},\dot{\boldsymbol{u}}^{(j)},\boldsymbol{x}(t)).
\end{equation}

In the next iteration, $\boldsymbol{\nu}^{(j)}$ is reallocated using $\mathbf{B}_s$ and the updated limits. The procedure continues until the required virtual control is achieved, no unsaturated actuators remain, or $N$ is reached. The final commanded control vector is
\begin{equation}
    \label{daisy_chain_u2}
    \boldsymbol{u}(t) = \sum_{j=1}^{N_{\text{iter}}} \boldsymbol{u}^{(j)}.
\end{equation}

For completeness, alongside the mathematical steps and the process visualization in Figure~\ref{fig:Control_Allocation_Process}, the algorithm is also summarized in pseudocode to complement the formulation and facilitate implementation.

\begin{algorithm}[ht]
\caption{Iterative Dynamic Control Allocation (IDCA)}
\begin{algorithmic}[1]
  \State \textbf{Initialize:} $N_{\text{iter}} \gets 1$, 
         $\mathbf{u}_{\max}^{(0)} \gets \mathbf{u}_{\max}(\mathbf{x},\bm{\sigma})$, 
         $\mathbf{u}_{\min}^{(0)} \gets \mathbf{u}_{\min}(\mathbf{x},\bm{\sigma})$,
         $\dot{\mathbf{u}}_{\max}^{(0)} \gets \dot{\mathbf{u}}_{\max}(\mathbf{x},\bm{\sigma})$,
         $\dot{\mathbf{u}}_{\min}^{(0)} \gets \dot{\mathbf{u}}_{\min}(\mathbf{x},\bm{\sigma})$,
         $\mathbf{u}_\tau \gets \mathbf{u}(t-T)$,
         $\mathbf{u}_s^{(0)} \gets \mathbf{u}_s$, 
         $\bm{\nu}^{(0)} \gets \bm{\nu}$
  \State Compute $\mathbf{u}_s^{(0)}, \mathbf{W}_m, \mathbf{W}_r$
  \While{$N_{\text{iter}} < N$}
      \State Compute candidate input $\mathbf{u}^{(j)}$ 
      \If{$\mathbf{u} = \sum_{j=1}^{N_{\text{iter}}} \mathbf{u}^{(j)}$ delivers desired $\bm{\nu}$}
          \State \textbf{Terminate with:} $\mathbf{u}(t) = \sum_{j=1}^{N_{\text{iter}}} \mathbf{u}^{(j)}$
      \EndIf
      \State Compute residuals of $\bm{\nu}^{(j)}$ and $\mathbf{u}_s^{(j)}$
      \If{$\mathrm{rank}(\mathbf{B}_s) = 0$}
          \State \textbf{Terminate with:} $\mathbf{u}(t) = \sum_{j=1}^{N_{\text{iter}}} \mathbf{u}^{(j)}$
      \EndIf
      \State Update actuator limits $\mathbf{u}_{\max}^{(j)}, \mathbf{u}_{\min}^{(j)}, \dot{\mathbf{u}}_{\max}^{(j)}, \dot{\mathbf{u}}_{\min}^{(j)}$
      \State Update iteration counter $N_{\text{iter}} \gets N_{\text{iter}} + 1$
  \EndWhile
  \State \textbf{Output:} final control input $\mathbf{u}(t)$
\end{algorithmic}
\end{algorithm}

\subsection{Control Input-Based Weighting Matrices}
\label{Actuator State-Based Weighting Matrices}
The main motivation of the here proposed iterative control allocation approach is to consider soft constraints on the distribution of the control efforts. To do so, a meaningful design of the weighting matrices $\mathbf{W}_m$ and $\mathbf{W}_r$ is leveraged. The magnitude weighting matrix $\mathbf{W}_m$ is designed to allocate weights to control effectors based on their current measured deflection state $u_i(t)$, as well as to use the weighting to prioritize control surfaces with less thermal loads acting on them. The overall matrix $\mathbf{W}_m$ is defined as:
\begin{equation}
 \mathbf{W}_m = 
 \begin{bmatrix}
w_{m,1}& 0 & 0 & 0\\ 
0 & w_{m,2}& 0 & 0\\ 
0 & 0 & w_{m,3}  & 0\\ 
0 & 0 & 0 & w_{m,4} 
\end{bmatrix} + \epsilon \mathbf{I},
\end{equation}
with $\epsilon \in \mathbb{R}_{>0}$ being a small positive scalar and $\mathbf{I}$ a square identity matrix used to prevent the loss of needed symmetric matrix properties in the case that one or more weight elements goes to zero. The elements $w_{m,i}(\boldsymbol{x}(t),u_i(t))$ represent, for each control surface $i$, a multiplicative weighting function designed to balance the two considered soft constraints in the allocation, as follows:
\begin{align}
    w_{m,i} = \underbrace{w_{m_D,i}(u_i(t))}_{\substack{\text{Weighting function} \\ \text{for deflection magnitude}}}  \cdot \underbrace{w_{m_T,i}(\boldsymbol{x}(t),\bm{\sigma}(t))}_{\substack{\text{Weighting function} \\ \text{for thermal loads}}},
\end{align} 
with $w_{m_D,i}(u_i(t))$ being the weighting function for the deflection magnitude and $w_{m_T,i}(\boldsymbol{x}(t),\bm{\sigma}(t))$ being the weighting function to address the effects of thermal loads for each effector $i$. The weighting function $w_{m_D,i}(u_i(t))$ is defined as:
\begin{equation}
w_{m_D,i}(u_i(t)) = \frac{u_i(t)}{u_{\max,i}(\boldsymbol{x}(t),\bm{\sigma}(t))}.
\end{equation}
{\color{black}
This formulation assigns higher weights to effectors with larger normalized deflection states, thereby discouraging further usage of actuators that are already operating close to their admissible limits. It is worth noting that any standard $\ell_2$-based control allocation, such as the Moore–Penrose pseudoinverse, inherently promote an even distribution of control effort. This follows directly from their objective of minimizing the Euclidean norm of the control input vector, even in the absence of explicit weighting. However, those unweighted formulations do not account for the instantaneous actuator state or for state-dependent variations in admissible control authority. As a result, actuators that are already highly deflected or close to saturation may continue to be utilized disproportionately, potentially leading to premature magnitude or rate saturation. By contrast, the proposed weighting explicitly incorporates the current actuator deflection relative to its state- and operating-point-dependent maximum admissible value. This enables a dynamic redistribution of control effort toward effectors with greater remaining authority, rather than enforcing uniform usage in an absolute sense. The effect becomes particularly important under asymmetric or state-dependent actuator limitations, such as those induced by varying aerodynamic and thermal loads in hypersonic flight, where different control surfaces may experience substantially different admissible ranges. In this way, the weighting extends the equalizing tendency of standard $\ell_2$-based allocation by making it aware of actuator availability.
}

\begin{figure}[h]
    \centering
    \includegraphics[width=0.9\textwidth]{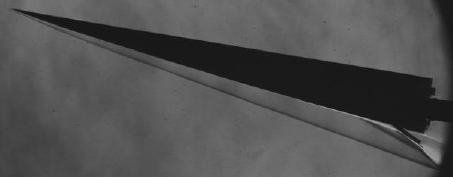}
    \caption{Wind tunnel test of GHGV-2 with \( M = 8.7 \), \( \alpha = 10^\circ \), and \( \beta = 0^\circ \); all flaps deflected by \( 20^\circ \)~\cite{Hohn2022}.}
    \label{fig:wind_tunnel_tests}
\end{figure}

To address the thermal effects via \( w_{m_T,i}(\boldsymbol{x}(t)) \) in a computationally efficient manner, we utilize insights from wind tunnel experiments conducted at the DLR Institute of Aerodynamics and Flow Technology~\cite{Hohn2022}. Figure~\ref{fig:wind_tunnel_tests} shows an image of the wind tunnel test of the GHGV-2 at Mach~8.7, angle of attack \( \alpha = 10^\circ \), and sideslip angle \( \beta = 0^\circ \), with all flaps deflected by \( 20^\circ \).

The obtained results demonstrated that during endoatmospheric flight at high Mach numbers, thermal loads on the vehicle's lower side are more critical than the upper side due to increased heat flux. This phenomenon is attributed to the vehicle's operation at predominantly positive angles of attack, which alters the aerodynamic heating distribution and results in less severe thermal conditions on the upper part of the vehicle. Therefore, it is preferable to utilize the upper flaps as much as possible during the mission while minimizing deflection of the lower flaps. The results also confirmed that thermal loads on the flaps increase with deflection angle for both upper and lower control surfaces. It is important to note that this distribution of thermal loads between upper and lower surfaces depends on the angle of attack: at positive angles of attack, the lower flaps experience higher thermal loads, whereas at negative angles of attack, the upper flaps are more critically affected. Similarly, sideslip angles can cause imbalances between the left and right sides of the vehicle.

Given the complex nature of thermal modeling, directly incorporating heat flux considerations into the computation of \( w_{m_T,i}(\boldsymbol{x}(t)) \) was deemed impractical. However, in all relevant cases, we observed a strong positive correlation between aerodynamic drag and heat flux with increasing control deflection angles across all control surfaces. The extent of this correlation varies depending on effector position. This correlation allows us to use aerodynamic drag as a proxy for thermal loads in the weighting function. Assuming these influences can be approximated as affine, the thermal weighting for control surfaces is computed as:
\begin{equation}
w_{m_T,i}(\boldsymbol{x}(t), \bm{\sigma}(t)) = \frac{C_{D,u_i}(\boldsymbol{x}(t))}{\max\{C_{D,u_1}(\boldsymbol{x}(t), \bm{\sigma}(t)),\hdots, C_{D,u_m}(\boldsymbol{x}(t), \bm{\sigma}(t))\}},
\end{equation}
where the denominator represents the maximum state-dependent aerodynamic drag coefficient among all control surfaces, and \( C_{D,u_i}(\boldsymbol{x}(t), \bm{\sigma}(t)) \) is the aerodynamic drag coefficient of control surface $i$. This formulation assigns higher weights to surfaces with higher drag contributions, thereby favoring allocation to surfaces with reduced thermal loads.

The rate weighting matrix $\mathbf{W}_r(\dot{\boldsymbol{u}}(t-T))$, designed to address the dynamic aspects of the control allocation problem by considering the rates of deflection, is defined as:
\begin{equation}
    \mathbf{W}_r = 
    \begin{bmatrix}
        w_{r,1} & 0 & 0 & 0 \\[6pt]
        0 & w_{r,2}& 0 & 0 \\[6pt]
        0 & 0 & w_{r,3} & 0 \\[6pt]
        0 & 0 & 0 & w_{r,4}
    \end{bmatrix} + \epsilon \mathbf{I},
\end{equation}
where \( w_{r,i} \) represents, for each control surface $i$, a rate-dependent weighting function defined as
\begin{equation}
    w_{r,i} = \frac{|\dot{u}_i(t)|}{\dot{u}_{\mathrm{crit}, i}(\boldsymbol{x}(t), \bm{\sigma}(t))},
\end{equation}
with the actuator rate at the initial time, when the allocation algorithm starts, given by
\begin{equation}
    \dot{u}_i(t) = \frac{1}{T} \big(u_i(t) - u_i(t-T)\big).
\end{equation}
The critical rate limit \(\dot{u}_{\mathrm{crit}, i}(\boldsymbol{x}(t))\), representing the rate bound the actuator is approaching, is defined as:
\begin{equation}
    \dot{u}_{\mathrm{crit}, i}(\boldsymbol{x}(t)) =
    \begin{cases}
        |\dot{u}_{\max,i}(\boldsymbol{x}(t), \bm{\sigma}(t))|, & \dot{u}_i(t) \geq 0, \\[6pt]
        |\dot{u}_{\min,i}(\boldsymbol{x}(t), \bm{\sigma}(t))|, & \dot{u}_i(t) < 0.
    \end{cases}
\end{equation}

The carefully crafted weighting matrices $\mathbf{W}_m$ and $\mathbf{W}_r$ contribute to the proposed control allocation algorithm's overall efficiency and stability, aligning with the discussed operational objectives. However, it is important to note that the proposed weighting functions $w_{m,i}$ and $w_{r,i}$ establish linear relationships with respect to the maximum and minimum values $u_{\max,i}(\boldsymbol{x}(t))$, $u_{\min,i}(\boldsymbol{x}(t))$, $\dot{u}_{\max,i}(\boldsymbol{x}(t))$, and $\dot{u}_{\min,i}(\boldsymbol{x}(t))$. Similarly, they assume an affine correlation between aerodynamic drag and thermal loads. While this linear approach ensures computational simplicity and efficiency, it inherently represents an idealized approximation. The proposed methodology simplifies the weighting function design but may not fully capture the nonlinear dependencies observed in practice.

\subsection{Designer Preference-Based Steady State Solution}
\label{Designer Preference-Based Steady State Solution}
The proposed IDCA approach allows it to consider a desired steady-state control input vector $\boldsymbol{u}_s(t)$. This section discusses how a good guess for $\boldsymbol{u}_s(t)$ is obtained based on an a priori known baseline control input $\boldsymbol{u}_r(t)$ and designer preferences in the control effector usage of the hypersonic glide vehicle. Even though the computed $\boldsymbol{u}_s(t)$ might not always provide a perfect solution to the requested virtual control command or in some cases might not be physically feasible under certain constraints, it serves as a desirable reference point or target for the redistributed control allocation algorithm. As discussed in Section~\ref{The Overall Control Allocation Concept}, in a scenario where no external disturbances or model uncertainties are present, the baseline control input $\boldsymbol{u}_r(t)$ would be sufficient to maintain the HGV on its reference trajectory. However, given the inherent uncertainties and potential disturbances in hypersonic flight, the control system also requires an attitude feedback controller to provide additional corrections. This controller computes a virtual control adjustment $\Delta \boldsymbol{\nu}(t)$, which compensates for deviations from the nominal trajectory. Assuming an affine nature for the control input, the desired steady state control input can be computed as:
\begin{equation}
    \boldsymbol{u}_s(t) = \boldsymbol{u}_r(t) + \Delta \boldsymbol{u}(t),
\end{equation}
where $\Delta \boldsymbol{u}(t)$ represents the correction determined through the proposed online control allocation algorithm. The control input $\Delta \boldsymbol{u}(t)$ is generated using the following request-dependent pseudoinverse-based approach:
\begin{align}
\Delta \boldsymbol{u}(t) = 
\begin{bmatrix} \Delta u_{1}(t) \\ \Delta u_{2}(t) \\ \Delta u_{3}(t) \\ \Delta u_{4}(t) \end{bmatrix}
= \mathbf{B}_{C}^\dagger(\Delta \boldsymbol{\nu}(t)) \, \Delta \boldsymbol{\nu}(t),
\quad \text{with} \quad
\mathbf{B}_{C}(\Delta \boldsymbol{\nu}(t)) =
\begin{bmatrix}
\mathbf{B}_{\nu_x}(\Delta \nu_x(t)) \\
\mathbf{B}_{\nu_y}(\Delta \nu_y(t)) \\
\mathbf{B}_{\nu_z}(\Delta \nu_z(t))
\end{bmatrix}.
\end{align}
Here $\mathbf{B}_{C}(\Delta \boldsymbol{\nu}(t))$ represents a conditionalized control effectiveness matrix. The submatrices $\mathbf{B}_{\nu_x}(\Delta \nu_x(t))$, $\mathbf{B}_{\nu_y}(\Delta \nu_y(t))$, and $\mathbf{B}_{\nu_z}(\Delta \nu_z(t))$ correspond to the three virtual control channels. The requested corrective virtual control command vector is
\[
\Delta \boldsymbol{\nu}(t) = [\Delta \nu_x(t), \Delta \nu_y(t), \Delta \nu_z(t)]^T,
\]
which provides information about the required virtual control moments along the body axes.\footnote{The abstract virtual controls $\nu_x, \nu_y, \nu_z$ correspond to the physical roll, pitch, and yaw moments $L, M, N$ in the body-fixed frame. We retain the abstract notation here to emphasize the generality of the formulation.} Based on flight dynamical properties and engineering preferences, certain elements of $\mathbf{B}_{\nu_x}(\Delta \nu_x(t))$, $\mathbf{B}_{\nu_y}(\Delta \nu_y(t))$, and $\mathbf{B}_{\nu_z}(\Delta \nu_z(t))$ are sparsified to force an allocation on desired control effectors. 

\begin{figure}[h!]
    \centering
    \includegraphics[width=0.7\linewidth]{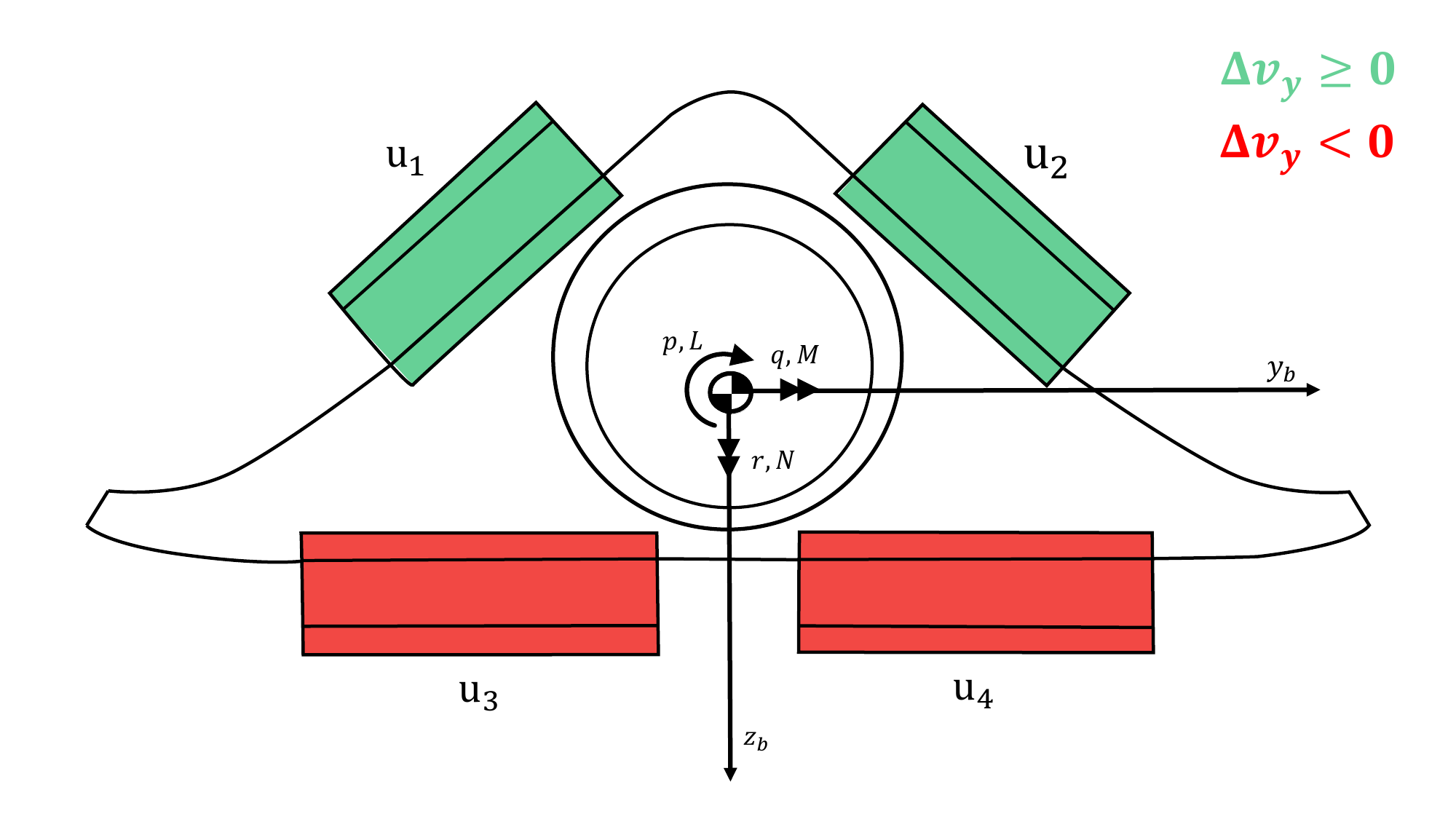}
    \caption{Corresponding control surface selection for positive and negative incremental pitch moment $\Delta \nu_y$ commands.}
\label{fig:ControlDeflectionSheme_Pitch}
\end{figure}

Figure~\ref{fig:ControlDeflectionSheme_Pitch} illustrates how the appropriate control surfaces are identified for an example of a virtual pitch command $\Delta \nu_y(t)$. For a positive pitch command $\Delta \nu_y(t) > 0$, both the upper flaps $u_{1}$ and $u_{2}$ are identified as appropriate. Conversely, for a negative pitch command $\Delta \nu_y(t) < 0$, $u_{3}$ and $u_{4}$ make the most sense from a flight dynamical perspective. This leads to the following sign-based conditionalization of $\mathbf{B}_{\nu_y}(\Delta \nu_y(t))$ in the pitch-related row:
\begin{equation}
\mathbf{B}_{\nu_y}(\Delta \nu_y(t)) = \begin{dcases*}
\begin{pmatrix} 
\displaystyle \frac{\partial \nu_y}{\partial u_{1}} & \displaystyle \frac{\partial \nu_y}{\partial u_{2}} & \displaystyle 0 & \displaystyle 0 
\end{pmatrix} & if $\Delta \nu_y(t) \geq 0$, \\[10pt]
\begin{pmatrix} 
\displaystyle 0 & \displaystyle 0 & \displaystyle \frac{\partial \nu_y}{\partial u_{3}} & \displaystyle \frac{\partial \nu_y}{\partial u_{4}} 
\end{pmatrix} & if $\Delta \nu_y(t) < 0$. \\
\end{dcases*}
\end{equation}

A similar approach is taken for the roll component $\Delta \nu_x(t)$ and the yaw component $\Delta \nu_z(t)$. The roll component $\mathbf{B}_{\nu_x}(\Delta \nu_x(t))$ is conditionalized in the following way:
\begin{equation}
\mathbf{B}_{\nu_x}(\Delta \nu_x(t)) = \begin{dcases*}
\begin{pmatrix} 
\displaystyle 0 & \displaystyle \frac{\partial \nu_x}{\partial u_{2}} & \displaystyle \frac{\partial \nu_x}{\partial u_{3}} & \displaystyle 0 
\end{pmatrix} & if $\Delta \nu_x(t) \geq 0$, \\[10pt]
\begin{pmatrix} 
\displaystyle \frac{\partial \nu_x}{\partial u_{1}} & \displaystyle 0 & \displaystyle 0 & \displaystyle \frac{\partial \nu_x}{\partial u_{4}} 
\end{pmatrix} & if $\Delta \nu_x(t) < 0$. \\
\end{dcases*}
\end{equation}
The yaw-related row $\mathbf{B}_{\nu_z}(\Delta \nu_z(t))$ is defined as:
\begin{equation}
\mathbf{B}_{\nu_z}(\Delta \nu_z(t)) = \begin{dcases*}
\begin{pmatrix} 
\displaystyle 0 & \displaystyle \frac{\partial \nu_z}{\partial u_{2}} & \displaystyle 0 & \displaystyle \frac{\partial \nu_z}{\partial u_{4}} 
\end{pmatrix} & if $\Delta \nu_z(t) \geq 0$, \\[10pt]
\begin{pmatrix} 
\displaystyle \frac{\partial \nu_z}{\partial u_{1}} & \displaystyle 0 & \displaystyle \frac{\partial \nu_z}{\partial u_{3}} & \displaystyle 0 
\end{pmatrix} & if $\Delta \nu_z(t) < 0$. \\
\end{dcases*}
\end{equation}
{\color{black}
The resulting control input $\boldsymbol{u}_s(t)$ constitutes a reasonable and operationally meaningful steady-state reference. It is important to emphasize that $\boldsymbol{u}_s(t)$ is neither enforced as a hard constraint nor required to exactly satisfy $\boldsymbol{\nu}(t) = \mathbf{B}\boldsymbol{u}(t)$. Instead, it serves as a designer-preferred initial target for the subsequent redistributed dynamic control allocation, as shown by the closed-form solution in Eq.~\eqref{linear_filter_solution}. The actual control input $\boldsymbol{u}(t)$ is obtained through an iterative redistribution process that explicitly accounts for feasibility, actuator limits, and the full control effectiveness matrix. If the reference $\boldsymbol{u}_s(t)$ is infeasible, the allocation naturally deviates from it and searches for a nearby admissible solution. From this perspective, the conditionalization can be interpreted as a structured initial guess that biases the solution toward desirable actuator usage patterns. Nevertheless, it must be noted that the induced sparsification implicitly introduces an allocation characteristic related to an $\ell_1$-type control allocation pattern. Such behavior is often advantageous in aerospace applications, as it promotes selective usage of the most effective control surfaces. However, as discussed in Section~\ref{The role of norms in control allocation}, purely $\ell_1$-based control allocation formulations tend to drive individual actuators toward their magnitude and rate limits, which can result in frequent constraint activation. In the proposed approach, this tendency is mitigated by embedding the designer-preferred steady-state solution as a soft reference within the proposed $\ell_2$-based dynamic control allocation approach. In this way, the sparsity-inducing properties of the steady-state preference are retained without being enforced as hard constraints, while the $\ell_2$-based dynamic control allocation law preserves compatibility with the full, non-simplified control effectiveness model as well as actuator magnitude and rate limitations.
}

%% file: Sections/Simulation_Results.tex
 A simulation-based analysis was carried out to examine the proposed IDCA algorithm. The flight dynamics model and the applied aerodynamic data set used for the time simulations are provided in a MATLAB/Simulink environment developed for the control design of hypersonic flight vehicles; see \cite{Autenrieb2021} for more details. 

\subsection{Comparison analysis on static moment commands}
\begin{figure}[ht]
\centering
\includegraphics[width=\columnwidth]{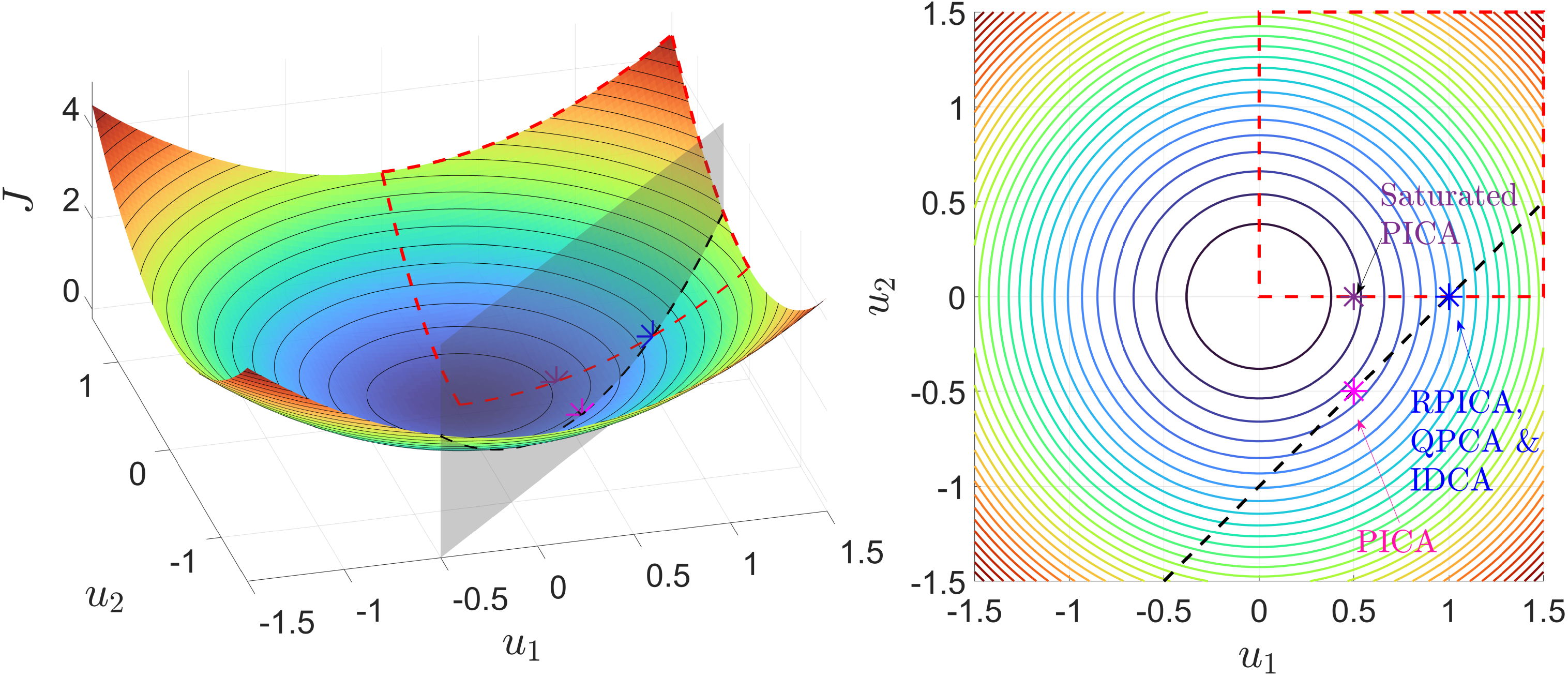}
\caption{Results of a simple constrained control allocation problem with two inputs and asymmetrical magnitude limits.}
\label{fig:Simulation example constrained control allocation}
\end{figure}

As discussed in Section~\ref{Control Allocation Algorithms}, the presented PICA and RPICA algorithms are not necessarily capable of handling asymmetrical control input limitations. To illustrate this limitation, a simplified example of a constrained control allocation problem is considered and visualized in Fig.~\ref{fig:Simulation example constrained control allocation}. The setup involves two control inputs \(\mathbf{u} = [u_1,\,u_2]^T\), where the admissible set is bounded by asymmetrical magnitude constraints \(\mathbf{u}_{\min} = [0,\,0]^T\) and \(\mathbf{u}_{\max} = [1.5,\,1.5]^T\). In this stationary example, rate constraints are neglected. The control effectiveness matrix is defined as \(\mathbf{B} = [0.5,\,-0.5]\), and the demanded virtual control input is given by \(\nu = 0.5\).

The set of all solutions to \(\nu = \mathbf{B}\mathbf{u}\) forms a straight line in the \((u_1,u_2)\)-plane, shown as the dashed black line in Fig.~\ref{fig:Simulation example constrained control allocation}. The admissible region for \(\mathbf{u}\) is marked by the dashed red box, representing the input bounds. Thus, the set of feasible solutions corresponds to the intersection between this line and the red box. Geometrically, every point on the black line segment that lies inside the red box is a feasible solution. Among these, the optimal solution is defined as the feasible point that minimizes the cost \(J = \|\mathbf{u}\|_2^2\), which corresponds to the point closest to the origin. In this case, the unique optimal solution is \(\mathbf{u} = [1,\,0]^T\).

As shown in the figure, the standard PICA algorithm computes the solution $\mathbf{u} = [0.5,-0.5]^T$ (magenta star), which satisfies the equation $\nu = \mathbf{B}\mathbf{u}$ but lies outside the admissible region and is therefore physically not feasible. After saturation, the corrected control input becomes $\mathbf{u} = [0.5,0]^T$ (violet star). This input is admissible but no longer reproduces the commanded virtual input, resulting in a large allocation error. By contrast, RPICA, QPCA, and the proposed IDCA algorithm are capable of incorporating the input limits directly and compute the admissible and optimal solution $\mathbf{u} = [1,0]^T$ (blue star). While RPICA is sufficient in this simple case, its performance deteriorates for higher-dimensional actuator systems with strong asymmetrical limits as present in the GHGV-2. QPCA is generally robust but computationally expensive, motivating the proposed IDCA algorithm that retains real-time feasibility while achieving comparable accuracy.

Based on this simple analysis, the algorithms are next evaluated at a representative operating point of the GHGV-2. The vehicle is trimmed for flight at Mach~$8$ and an altitude of $30$\,km, where the four aerodynamic control surfaces of the HGV $u_1$ to $u_4$ are available. The admissible bounds are defined as
\[
\mathbf{u}_{\min} = 
\begin{bmatrix}
0^\circ & 0^\circ & 0^\circ & 0^\circ
\end{bmatrix}^T,
\qquad
\mathbf{u}_{\max} = 
\begin{bmatrix}
20^\circ & 20^\circ & 20^\circ & 20^\circ
\end{bmatrix}^T,
\]
and the control effectiveness matrix at this operating point is
\begin{equation}
\label{eqn:Control_effectiveness_matrix}
\mathbf{B} = 
\begin{pmatrix} 
-20.01 & 20.01 & 93.94 & -93.94 \\[8pt]
126.7 & 126.7 & -501.4 & -501.4 \\[8pt]
-127.5 & 127.5 & -45.72 & 46.72
\end{pmatrix}.
\end{equation}
A stationary virtual control command \(\nu = [-400,\,800,\,-2000]^T\)~Nm is prescribed, which lies within the AMS and is therefore physically realizable without violating the actuator magnitude limits. Since the command is stationary, rate constraints are not considered in this case.

\begin{figure}
\centering
\includegraphics[width=0.7\columnwidth]{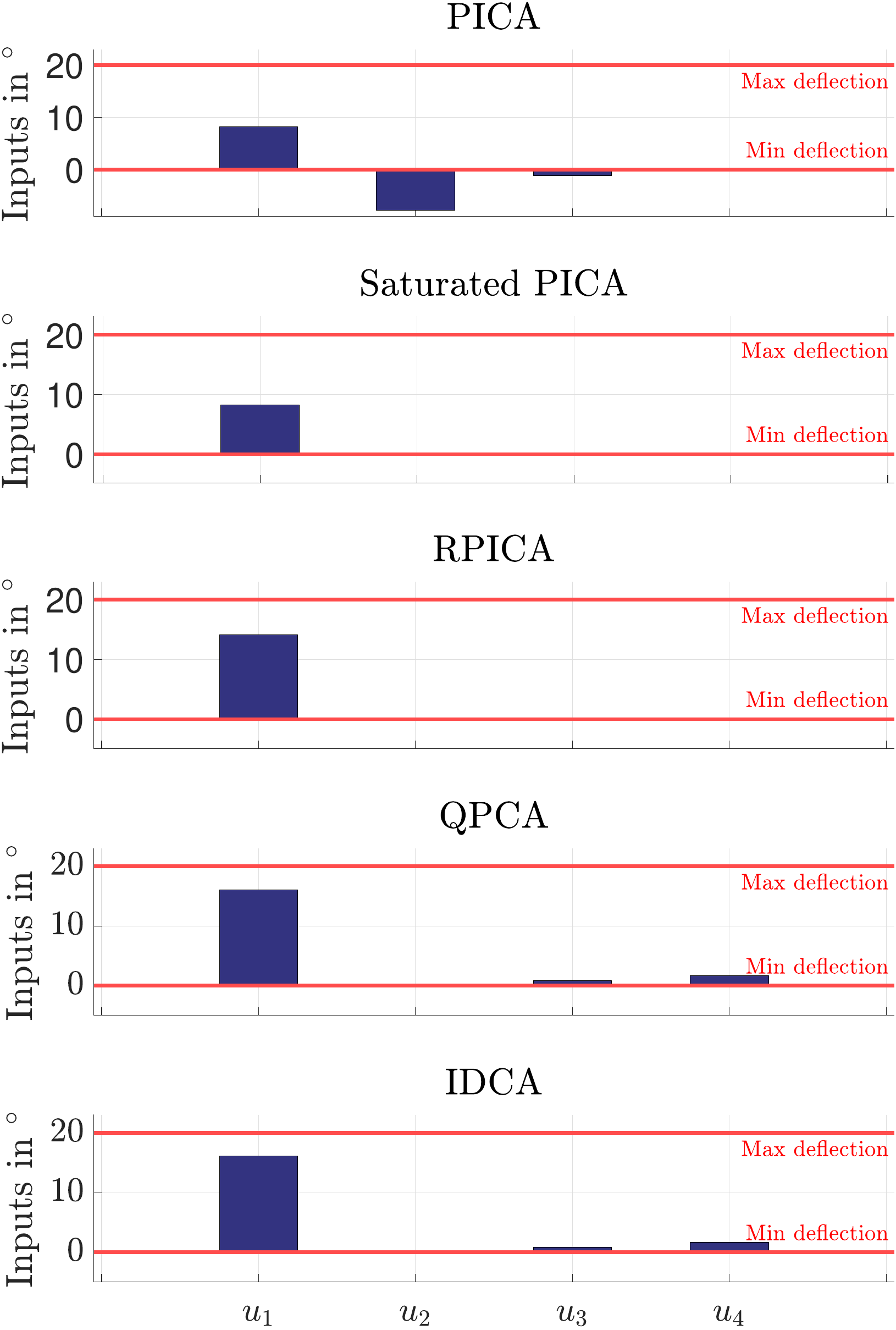}
\caption{Obtained control deflections for different allocation algorithms under a moment command within the AMS of the GHGV-2.}
\label{fig:table_no_saturation}
\end{figure}

The results are shown in Fig.~\ref{fig:table_no_saturation}. PICA generates a solution that formally satisfies the allocation equation but violates the lower input bounds for several surfaces, making it physically infeasible. Saturated PICA respects the limits but concentrates almost the entire effort on $u_1$, failing to reproduce the commanded moment. RPICA also respects the bounds but suffers from premature saturation in three surfaces during the first iteration, leaving only $u_1$ to carry the residual, which again results in a large allocation error. In contrast, QPCA and IDCA both provide admissible deflections that satisfy the command. Their allocations are nearly identical, with QPCA solving a quadratic optimization problem and IDCA iterating dynamically over weighted residuals.

\begin{table}[hbt!]
\caption{\label{tab:table_no_saturation} Results for the GHGV-2 stationary allocation case.}
\centering
\begin{tabular}{lccc}
\hline
\textbf{Method} & \textbf{Cost $J=\|\mathbf{u}\|_2$ [Nm]} & \textbf{Error norm $\|\nu-\mathbf{B}\mathbf{u}\|_2$ [Nm]} & \textbf{Computation time [s]} \\
\hline
PICA & $11.3749$ & $1.3270\times 10^{-12}$ & $1.5760 \times 10^{-4}$ \\
Saturated PICA & $8.1773$ & $1.0140\times 10^{3}$ & $1.5760 \times 10^{-4}$ \\
RPICA & $14.1015$ & $1.0140\times 10^{3}$ & $3.9210 \times 10^{-4}$ \\
QPCA & $16.1473$ & $2.2737\times 10^{-13}$ & $0.3911$ \\
IDCA & $16.2146$ & $3.2155\times 10^{-13}$ & $5.5550 \times 10^{-4}$ \\
\hline
\end{tabular}
\end{table}

Table~\ref{tab:table_no_saturation} provides the allocation cost \(J=\|\mathbf{u}\|_2\), the allocation error \(\|\nu-\mathbf{B}\mathbf{u}\|_2\), and the computation time. PICA exhibits the lowest cost but yields an invalid solution, while saturated PICA and RPICA produce high allocation errors despite respecting the bounds. QPCA and IDCA achieve nearly identical, minimal error norms, but QPCA requires several orders of magnitude longer computation time. IDCA achieves the same level of accuracy at real-time-capable runtimes, only marginally slower than PICA and RPICA.  

To further assess robustness, a Monte Carlo analysis with \(N=1000\) random virtual control commands is conducted. The mean command vector is chosen as \(\nu_{\mathrm{m}} = [-100,\,300,\,-500]^T\), with each component modeled as a Gaussian random variable \(\nu_i \sim \mathcal{N}(\nu_{\mathrm{m},i},\sigma_i^2)\). The standard deviations are set to \(\sigma_i = \tfrac{2}{3}|\nu_{\mathrm{m},i}|\) for \(i=1,2,3\), such that the \(\pm 3\sigma\) range spans approximately twice the mean value of each component. Random commands are then generated according to \(\nu = \nu_{\mathrm{m}} + \boldsymbol{\sigma} \odot \mathbf{z}\), where \(\mathbf{z}\) is a vector of independent standard normal variables and \(\odot\) denotes element-wise multiplication. A total of \(N=1000\) such samples is generated, ensuring a broad coverage of possible commands both near and within the AMS. The actuator limits and the control effectiveness matrix remain unchanged from the stationary case.

\begin{figure}[h!]
\centering
\includegraphics[width=0.5\columnwidth]{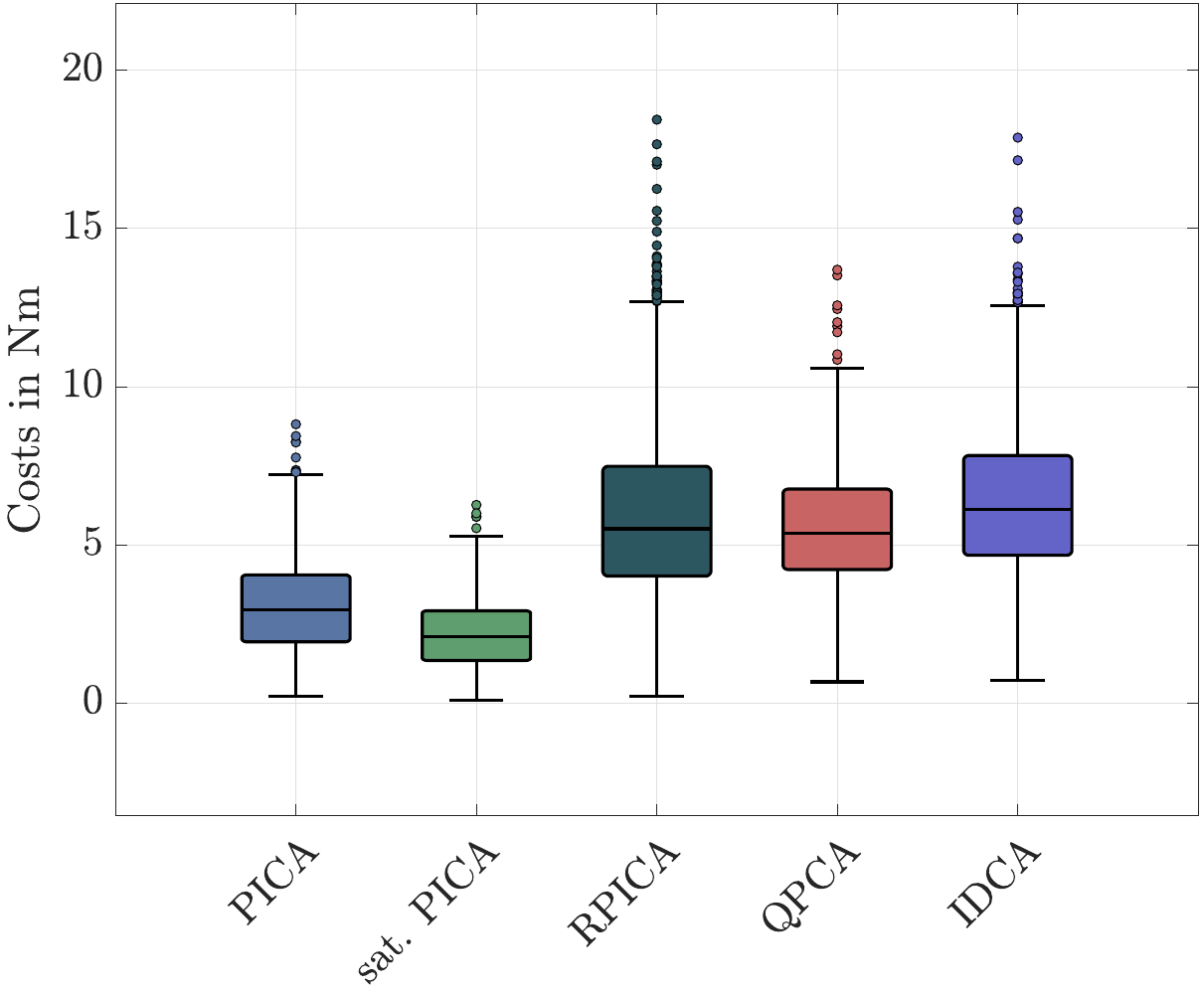}
\caption{Distribution of allocation costs $J=\|\mathbf{u}\|_2$ over $N=1000$ Monte Carlo samples.}
\label{fig:Monte_Carlo_Kosten}
\end{figure}

Figure~\ref{fig:Monte_Carlo_Kosten} shows the cost distributions. IDCA produces slightly higher costs than QPCA and RPICA on average, but all results remain within physically reasonable ranges. PICA frequently exhibits lower costs, though many of its solutions are inadmissible.

\begin{figure}[h!]
\centering
\includegraphics[width=0.5\columnwidth]{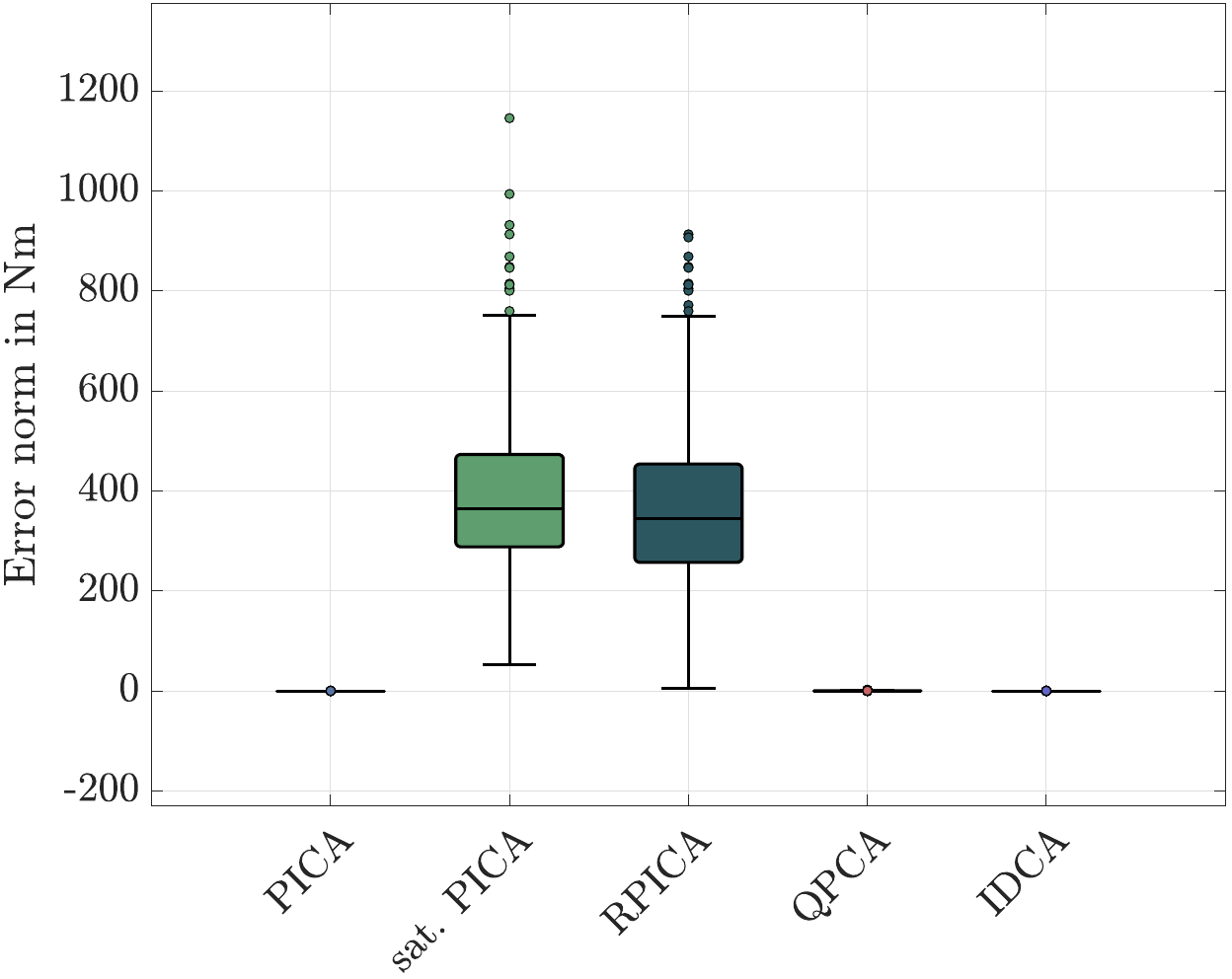}
\caption{Distribution of allocation error $\|\nu - \mathbf{B}\mathbf{u}\|_2$ over $N=1000$ Monte Carlo samples.}
\label{fig:Monte_Carlo_Fehler}
\end{figure}

The corresponding allocation error distributions in Fig.~\ref{fig:Monte_Carlo_Fehler} reveal that PICA, QPCA, and IDCA achieve consistently low errors, whereas saturated PICA and RPICA produce significantly larger errors, often exceeding several hundred Nm. This confirms that PICA neglects input limits, while QPCA and IDCA generate realizable allocations.

\begin{figure}[h!]
\centering
\includegraphics[width=0.5\columnwidth]{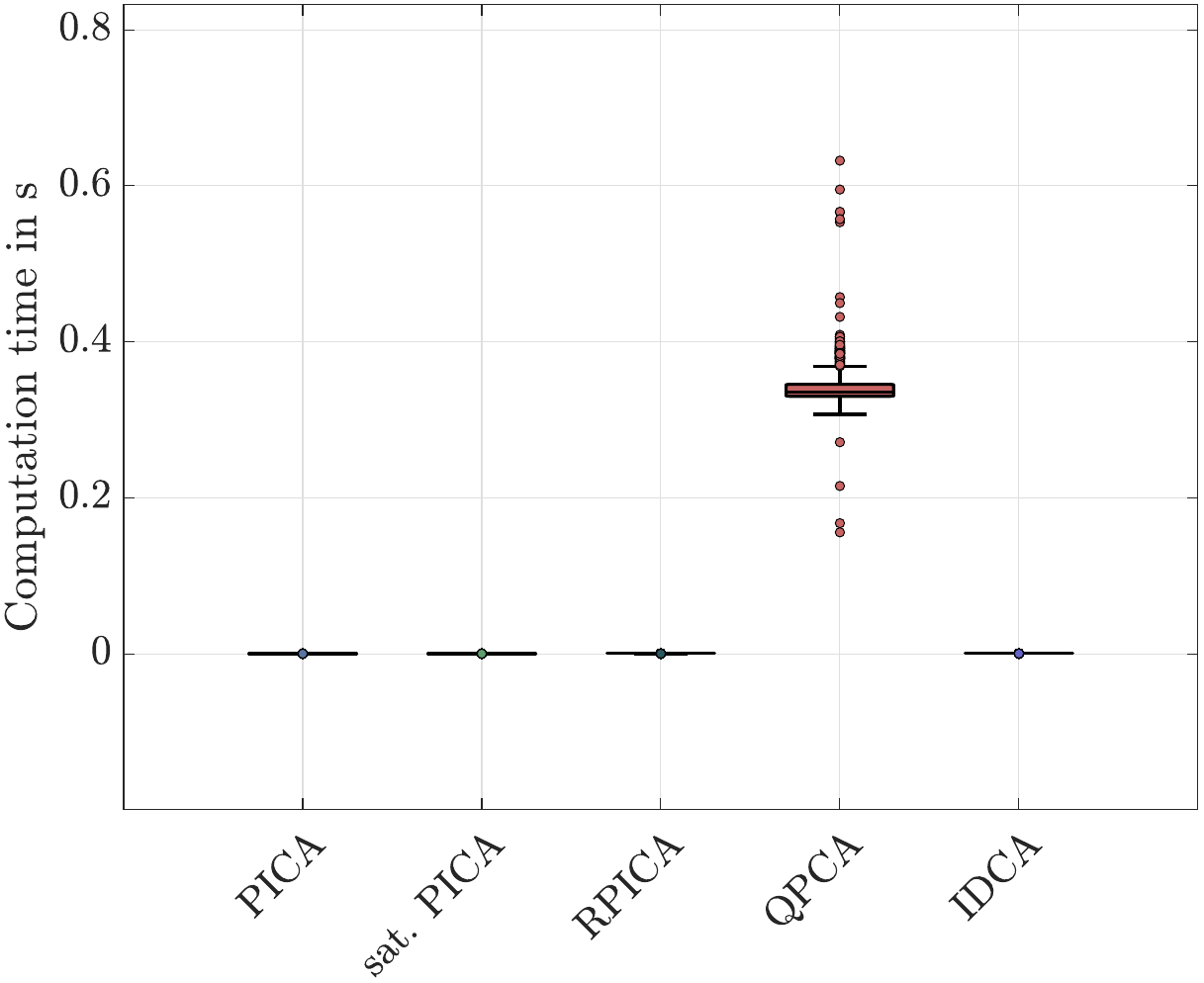}
\caption{Computation times across $N=1000$ Monte Carlo samples.}
\label{fig:Monte_Carlo_Zeit}
\end{figure}

Finally, Fig.~\ref{fig:Monte_Carlo_Zeit} compares computation times. QPCA requires orders of magnitude longer runtimes due to solving a quadratic program at every iteration, while PICA, saturated PICA, and RPICA remain computationally fast but unreliable. IDCA achieves runtimes in the millisecond range, slightly above RPICA, but still compatible with real-time requirements.

Overall, the stationary and Monte Carlo analyses confirm that while PICA produces formally correct but inadmissible solutions, and RPICA respects constraints but distributes residuals poorly, QPCA and IDCA achieve accurate, realizable allocations. The decisive advantage of IDCA is that it matches the accuracy of QPCA at real-time-capable runtimes, demonstrating its suitability for onboard implementation in the GHGV-2.

\subsection{Time-varying moment commands on proposed algorithm}
For this assessment, the same control effectiveness matrix \( \mathbf{B} \) as in Eq.~\eqref{eqn:Control_effectiveness_matrix} is considered. The GHGV-2 is again trimmed at Mach~8 and an altitude of 30\,km, representing the same operating condition as before. In this analysis, however, the magnitude limits of the control surfaces are modeled as time-varying functions in order to capture the dynamic effects of varying aerodynamic loads at high Mach numbers. It is assumed that the influence of dynamic pressure leads to a reduction in the achievable upper magnitude bound of the actuators, while the lower magnitude bound remains at zero for all control surfaces, as it is not significantly affected by external influences due to the direction of the resulting aerodynamic forces. To model this effect, the upper bounds are expressed as a sinusoidal modulation:
\begin{equation*}
u_{\text{max}}(t) = u_{\text{max,full}} \cos(\Lambda(t)),
\end{equation*}
where \(u_{\text{max,full}}=20^\circ\) denotes the technically achievable maximum flap deflection and \(\Lambda(t)\) is a freely chosen modulation function.  

In addition to time-varying magnitude bounds, rate constraints are considered as well. The upper and lower bounds of the rates, \( |\dot{\boldsymbol{u}}_{\max}| \) and \( |\dot{\boldsymbol{u}}_{\min}| \), are modeled as time-dependent functions. The upper rate limit decreases from \(20^\circ/\text{s}\) to \(10^\circ/\text{s}\) over the simulation, reflecting the reduced ability of actuators to achieve high positive rates under increasing aerodynamic loads. Conversely, the lower rate limit is assumed to increase from \(-20^\circ/\text{s}\) to \(-30^\circ/\text{s}\), because the aerodynamic forces facilitate the retraction of the control surfaces. This asymmetry captures the physical reality that positive rate changes become more restricted, while negative rate changes become easier to achieve as the flight condition evolves.

\begin{figure}
	\centering
	\includegraphics[width=\columnwidth]{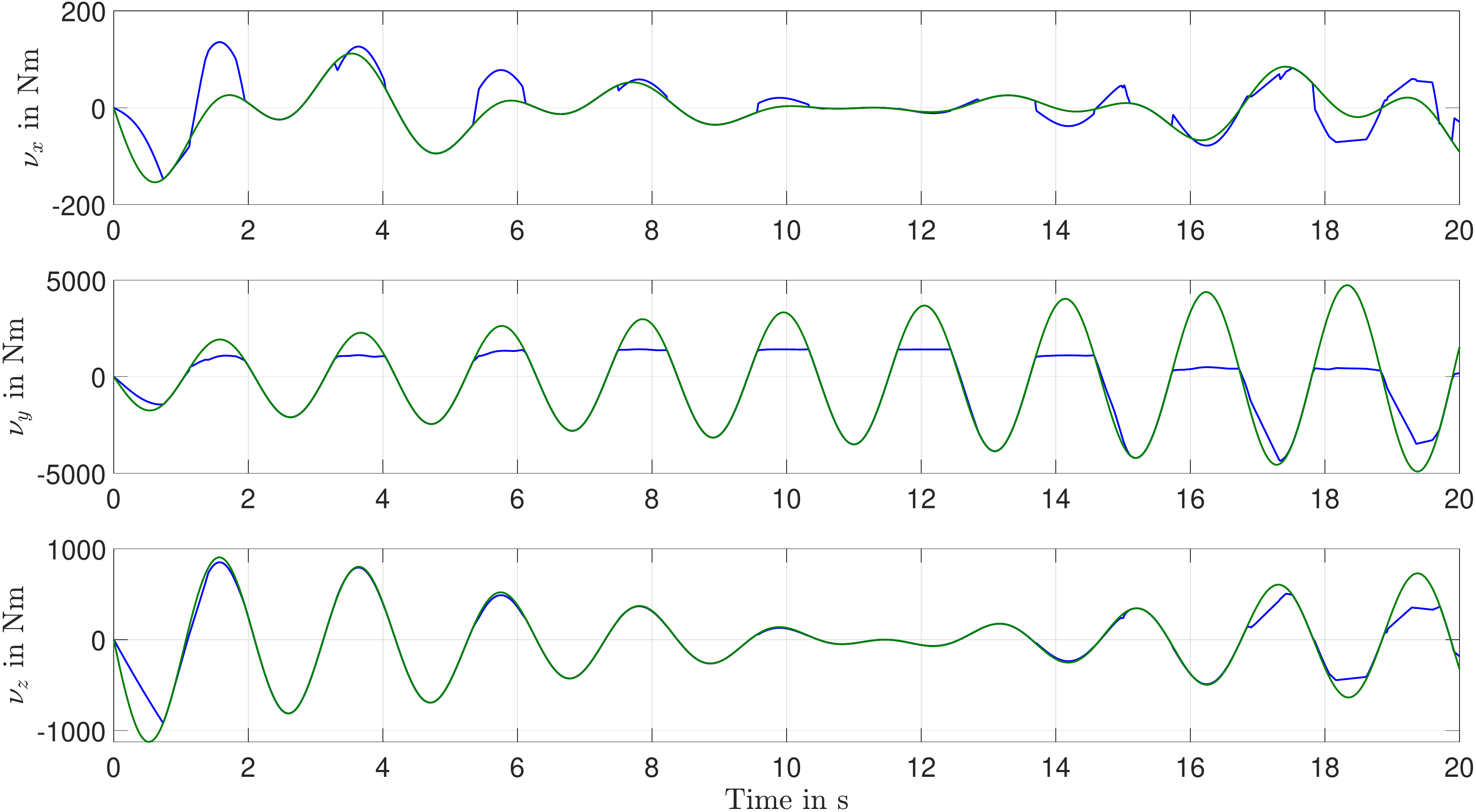}
	\caption{Results for the virtual control commands \(\nu_x\), \(\nu_y\), and \(\nu_z\). Green: commanded input. Blue: generated input after allocation.}
	\label{fig:Moment_Results}
\end{figure}

The allocation module is tested with a sinusoidal virtual control input command \(\bm\nu(t) = [\nu_x(t), \nu_y(t), \nu_z(t)]^T\). The chosen command alternates between feasible and infeasible values with respect to the AMS, enabling an evaluation of the algorithm’s performance under both realizable and saturated conditions. Figure~\ref{fig:Moment_Results} compares the commanded virtual control inputs (green) with the moments generated by the IDCA algorithm (blue). A generally close tracking is observed, particularly for \(\nu_z\), while noticeable deviations appear in \(\nu_x\) and \(\nu_y\) whenever actuator magnitude or rate saturation is encountered. During unsaturated phases the algorithm achieves high accuracy, with nearly perfect matching of the requested commands.

\begin{figure}
	\centering
	\includegraphics[width=\columnwidth]{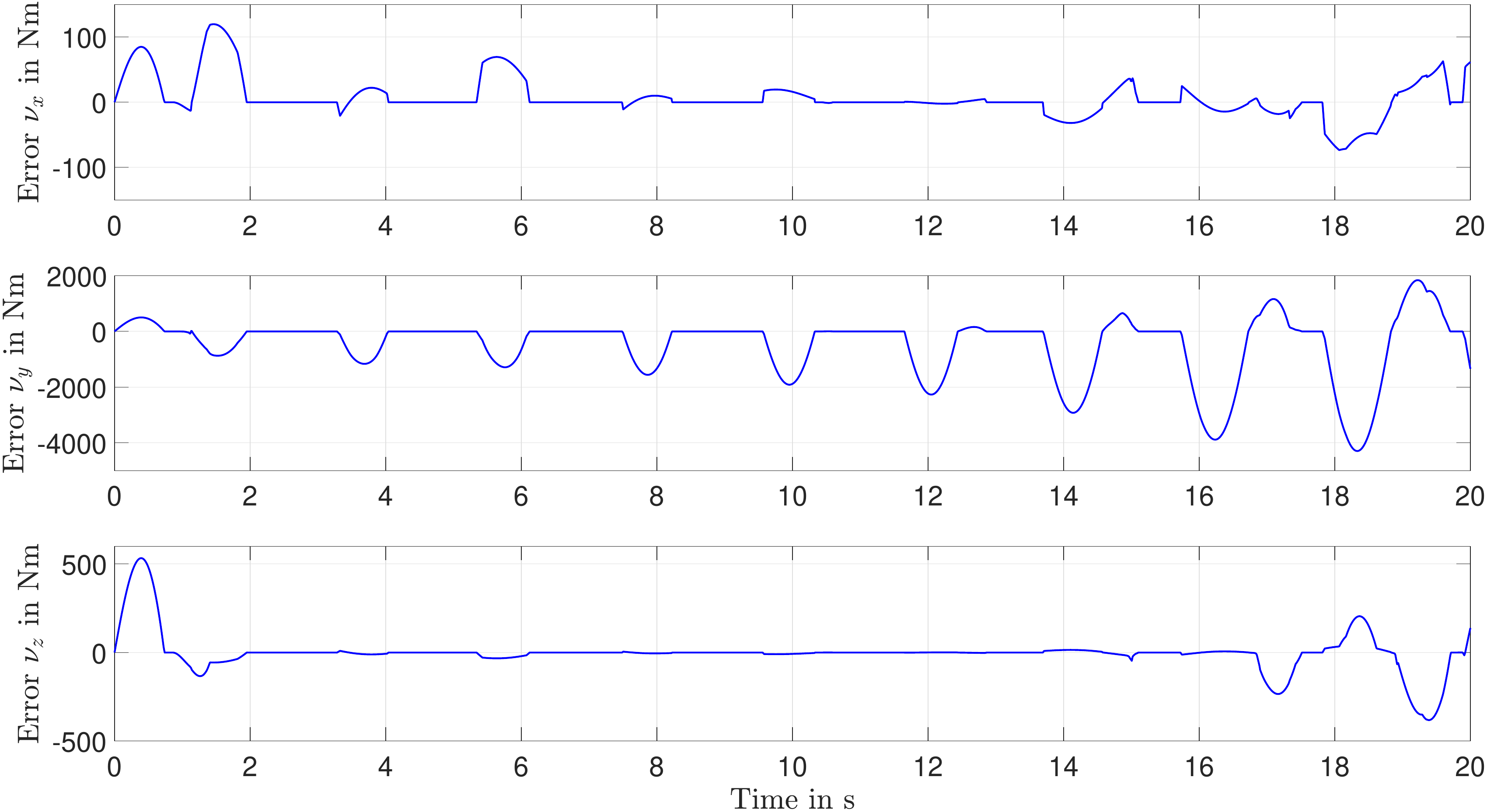}
	\caption{Error between commanded and generated virtual control inputs \(\nu_x\), \(\nu_y\), and \(\nu_z\). Blue lines: allocation error time series.}
	\label{fig:Error_Results}
\end{figure}

Figure~\ref{fig:Error_Results} shows the error between the requested and generated commands, defined as
\[
\Delta \bm\nu(t) = \bm\nu(t) - \mathbf{B}\boldsymbol{u}(t).
\]
The results confirm that the error converges towards zero in non-saturated phases, while in saturated regimes distinct deviations occur. The error magnitudes align with the observations from Fig.~\ref{fig:Moment_Results}, where \(\nu_y\) exhibits the largest deviations during phases of strong demand. This correlation indicates that actuator constraints directly determine the achievable accuracy of the allocation.

\begin{figure}
	\centering
	\includegraphics[width=\columnwidth]{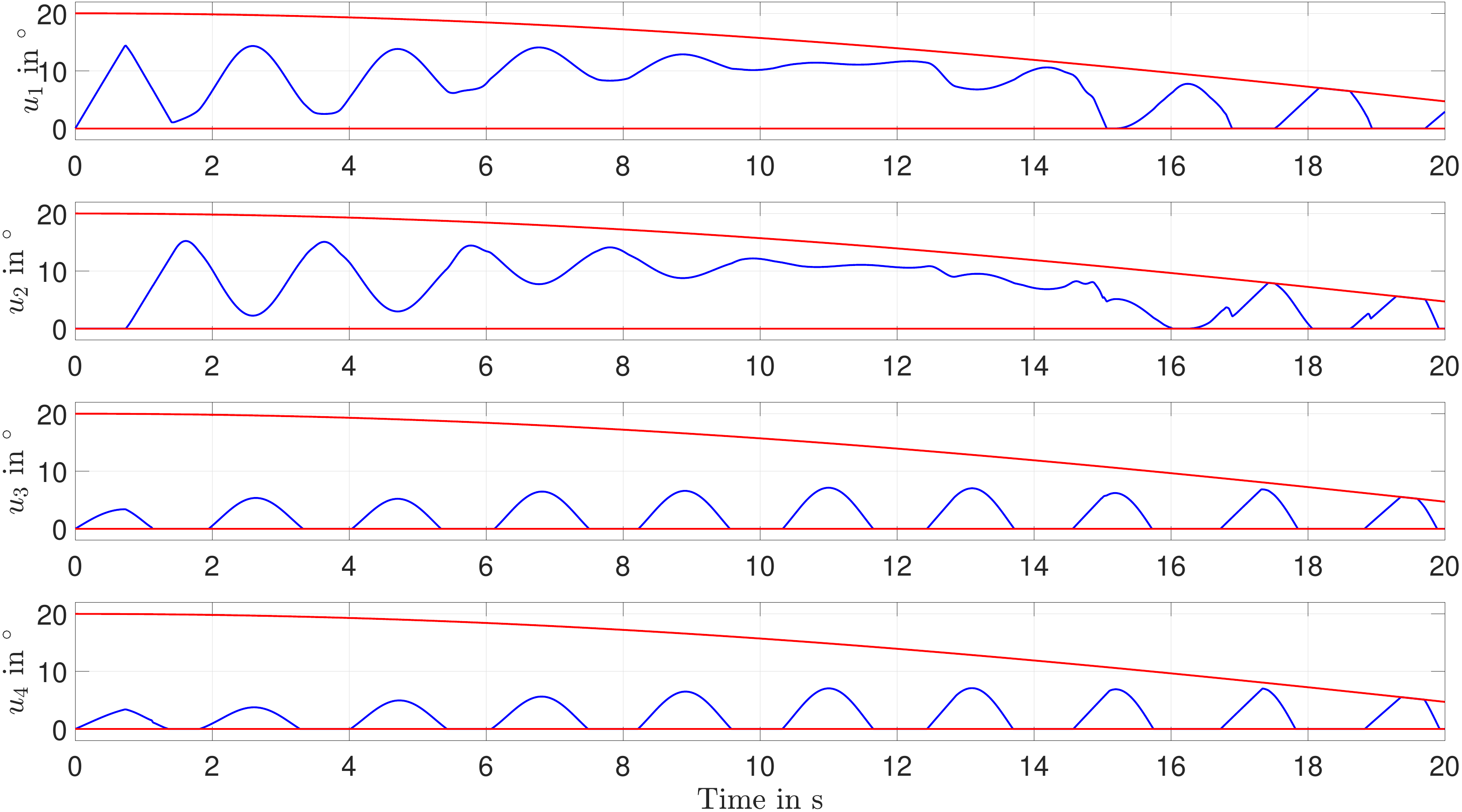}
	\caption{Computed control input magnitudes with time-varying magnitude constraints. Blue: generated deflections. Red: admissible bounds.}
	\label{fig:Magnitude_Results}
\end{figure}

\begin{figure}
  \centering
  \includegraphics[width=\columnwidth]{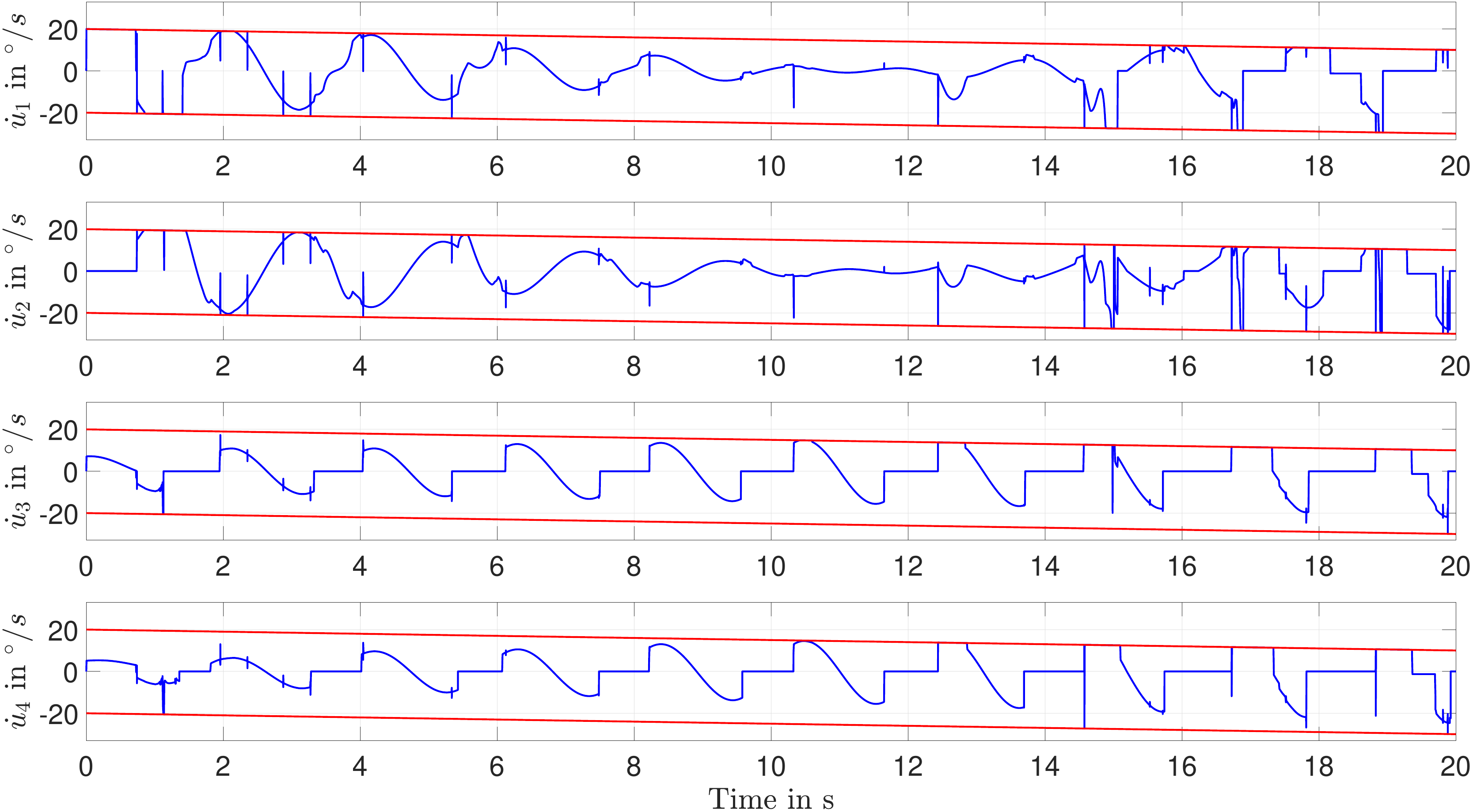}
  \caption{Computed control input rates with time-varying rate constraints. Blue: generated rates. Red: admissible bounds.}
  \label{fig:Rate_Results}
\end{figure}

The actuator magnitudes \(u_1,\dots,u_4\) and their time-varying bounds are shown in Fig.~\ref{fig:Magnitude_Results}. The trajectories clearly demonstrate that all commands remain within the admissible set. In the initial phase, a nearly linear increase is visible for \(u_1\) and \(u_2\), which indicates active rate saturation: since the maximum rate is reached, the magnitude grows linearly until the limit changes. Comparing with Fig.~\ref{fig:Rate_Results} confirms this behavior, as \(\dot u_1\) and \(\dot u_2\) remain pinned at their respective limits during these phases. The corresponding deviations in \(\Delta \nu_x\) and \(\Delta \nu_y\) highlight the direct causal link between actuator rate constraints and tracking error.

The role of the control surfaces in producing the commanded virtual inputs is also evident. For example, to generate positive \(\nu_y\), the algorithm retracts the surfaces that would produce negative contributions (\(u_3, u_4\)) while driving \(u_1, u_2\) upward. Conversely, for negative \(\nu_y\), the roles switch. A similar logic applies for roll moments \(\nu_x\), where surfaces \(u_2, u_3\) dominate the positive direction, while \(u_1, u_4\) dominate the negative. This role distribution is consistent with the control effectiveness structure in Eq.~\eqref{eqn:Control_effectiveness_matrix} and explains why certain surfaces repeatedly saturate depending on the sign of the demanded moment. The algorithm implicitly prioritizes the dominant axis when strong demands occur, reducing error in that channel even if moderate errors appear in the others. For instance, in phases of high pitch demand, the allocation accepts deviations in \(\nu_x\) to minimize the error in \(\nu_y\). Such behavior could be adapted with weighting factors to impose mission-specific priorities, though in the present setup the distribution already provides a reasonable compromise.

The time histories of the rates in Fig.~\ref{fig:Rate_Results} further illustrate this behavior. Whenever a magnitude limit is reached, the corresponding rate collapses to zero, consistent with the physical limit of the actuator. This produces linear evolution in the magnitude traces and explains the sharp transitions observed in the error plots. Spikes in the rate signals occur when one surface saturates and the remaining effort is redistributed across the other actuators, causing rapid but coordinated adjustments. These interactions confirm the coupled nature of the allocation: constraints in one actuator channel propagate into others as the residual command is redistributed.

Overall, the simulation results present a consistent and physically plausible picture. The time-varying magnitude and rate bounds are respected at all times, the algorithm provides accurate tracking in feasible regions, and deviations are directly attributable to actuator limits. The prioritization behavior ensures that dominant demands are satisfied first, yielding a robust trade-off across channels. Importantly, even under dynamic and asymmetric constraints, the IDCA algorithm demonstrates reliable real-time operation and robust allocation, confirming its applicability for hypersonic flight scenarios where actuator characteristics vary significantly over time.

%% file: Sections/Conclusions.tex
{\color{black}
In conclusion, this paper presented an iterative dynamic control allocation approach tailored to hypersonic glide vehicles and evaluated it in the context of asymmetric input limitations, actuator coupling, and state-dependent operating conditions. In contrast to pseudoinverse-based and redistributed pseudoinverse methods, the proposed formulation explicitly accounts for asymmetric magnitude and rate constraints and avoids degenerate or infeasible solutions under strong actuator coupling. Compared to quadratic programming–based control allocation approaches, the method significantly reduces computational complexity while retaining the ability to shape control distribution through soft constraints, making it suitable for real-time implementation in inner-loop flight control systems.

A key aspect of the proposed approach lies in the formulation of the control allocation problem itself. By appropriately designing state- and actuator-dependent weighting functions, the allocation process accounts for the instantaneous magnitude and rate states of the actuators as well as their directional control effectiveness. This enables a more balanced distribution of control effort and reduces the tendency of individual actuators to saturate, thereby preserving control authority under constrained conditions. In addition, the weighting function design provides a systematic mechanism to address thermal load effects. The use of drag-sensitive weighting functions allows thermal considerations to be incorporated implicitly into the allocation process without introducing additional hard constraints. By influencing control distribution based on drag generation characteristics, thermal loads and associated energy losses can be reduced while maintaining feasibility and compatibility with an iterative, approximate solution strategy. 

Simulation results using the GHGV-2 model demonstrate that the proposed approach reliably respects actuator magnitude and rate limits, maintains control authority under realistic flight conditions, and offers a practical and computationally efficient alternative to existing state-of-the-art control allocation techniques for hypersonic applications.
}
